\documentclass[twocolumn,tighten]{aastex62}
\usepackage{wasysym}

\newcommand{\sion}[2]{#1$\;$\textsc{#2}\relax}
\newcommand{\sub}[2]{\ifmmode #1_\mathrm{\scriptstyle #2} \else $#1_\mathrm{\scriptstyle #2}$\fi}
\newcommand{\ssub}[2]{\ifmmode #1_\mathrm{\scriptscriptstyle #2} \else $#1_\mathrm{\scriptscriptstyle #2}$\fi}
\newcommand{\pme}[2]{$^{+#1}_{-#2}$}
\newcommand{\pe}[1]{$^{+#1}$}
\newcommand{\me}[1]{$_{-#1}$}
\shorttitle{\textit{HST}/COS Observations of Outflows in PKS J0352-0711}
\shortauthors{Miller et al.}
\begin{document}

\title{\textit{HST}/COS Observations of Quasar Outflows in the 500 -- 1050 \AA\ Rest Frame: V\\Richness of Physical Diagnostics and Ionization Potential-dependent Velocity Shift in PKS J0352-0711}
\altaffiliation{Based on observations with the NASA/ESA \\\textit{Hubble Space Telescope} obtained at the Space Telescope \\Science Institute, which is operated by the Association \\of Universities for Research in Astronomy, Incorporated, \\under NASA contract NAS5-26555.}

\author[0000-0002-0730-2322]{Timothy R. Miller}
%\email{tim@huskers.unl.edu}
\affiliation{Department of Physics, Virginia Polytechnic Institute and State University, Blacksburg, VA 24061, USA}
\author[0000-0003-2991-4618]{Nahum Arav}
\affiliation{Department of Physics, Virginia Polytechnic Institute and State University, Blacksburg, VA 24061, USA}
\author[0000-0002-9217-7051]{Xinfeng Xu}
\affiliation{Department of Physics, Virginia Polytechnic Institute and State University, Blacksburg, VA 24061, USA}
\author[0000-0002-2180-8266]{Gerard A. Kriss}
\affiliation{Space Telescope Science Institute, 3700 San Martin Drive, Baltimore, MD 21218, USA}
\author{Rachel J. Plesha}
\affiliation{Space Telescope Science Institute, 3700 San Martin Drive, Baltimore, MD 21218, USA}
%\author{Kirk T. Korista}
%\affiliation{Department of Physics, Western Michigan University, Kalamazoo, MI 49008-5252, USA}
%\author{Chris Benn}
%\affiliation{Isaac Newton Group, Apartado 321, E-38700 Santa Cruz del La Palma, Spain}
%\author{Guilin Liu}
%\affiliation{CAS Key Laboratory for Research in Galaxies and Cosmology, Department of Astronomy, University of Science and Technology of China, Hefei 230026, China}
%\affiliation{School of Astronomy and Space Sciences, University of Science and Technology of China, Hefei 230026, China}

%% Note that the \and command from previous versions of AASTeX is now
%% depreciated in this version as it is no longer necessary. AASTeX 
%% automatically takes care of all commas and "and"s between authors names.

%% AASTeX 6.2 has the new \collaboration and \nocollaboration commands to
%% provide the collaboration status of a group of authors. These commands 
%% can be used either before or after the list of corresponding authors. The
%% argument for \collaboration is the collaboration identifier. Authors are
%% encouraged to surround collaboration identifiers with ()s. The 
%% \nocollaboration command takes no argument and exists to indicate that
%% the nearby authors are not part of surrounding collaborations.

%% Mark off the abstract in the ``abstract'' environment. 
\begin{abstract}
We analyze absorption troughs from two outflows within PKS J0352-0711 (S1 at $-$1950~km~s$^{-1}$ and S2 at $-$3150~km~s$^{-1}$) from spectra taken with \textit{Hubble Space Telescope}/Cosmic Origin Spectrograph, which cover the diagnostic-rich 585-900~\AA\ rest frame wavelength range. In S2, for the first time we clearly detect absorption troughs from \sion{Ca}{iv}, \sion{Ca}{v}, \sion{Ca}{v*}, \sion{Ca}{vii*}, and \sion{Ca}{viii*}. The column density measurement of \sion{Ca}{v} suggests S2 has a super-solar metallicity. Both outflows require at least two ionization phases where the column density of the very high-ionization phase is roughly 15 times larger than the corresponding high-ionization phase. These high column densities and very high-ionization potential ions are similar to X-ray warm absorbers. The two phases of S2 show a unique velocity centroid shift between associated troughs. Through Monte Carlo measurements of the \sion{O}{v*} absorption troughs, we determine the electron number density of S2 (fully corroborated by independent measurements from the \sion{Ca}{vii*} and \sion{Ca}{viii*} troughs), yielding a distance of 9~pc and a kinetic luminosity of 2$\times$10$^{43}$~erg~s$^{-1}$. S1 is located farther away at 500~pc from the central source with a kinetic luminosity of 10$^{43}$~erg~s$^{-1}$. 
\end{abstract}
%% Keywords should appear after the \end{abstract} command. 
%% See the online documentation for the full list of available subject
%% keywords and the rules for their use.
\keywords{galaxies: active --- galaxies: kinematics and dynamics --- ISM: jets and outflows --- quasars: absorption lines --- quasars: general --- quasars: individual(PKS J0352-0711)}

\section{Introduction}\label{sec:int}
Quasar outflow systems are typically identified from blueshifted absorption troughs observed in the rest frame of quasar spectra. These outflows are found in 20-40\% of the quasar population \cite[e.g.,][]{hew03,dai08,gan08,kni08}. The distance these outflows are from the central source ($R$) can be inferred from simultaneously determining the electron number density (\sub{n}{e}) and ionization parameter (\sub{U}{H}) of the outflow \cite[e.g.,][]{ara13}. To date, around 20 such distances have been published by our group and others using this method (see section 1 of \cite{ara20a}, hereafter Paper I, and references therein). These distances are in the range of parsecs to tens of kiloparsecs, orders of magnitude more distant than accretion disk wind models predict \cite[$\sim$0.03~pc; e.g.,][]{mur95,pro00,pro04}.
%\\\\These outflow systems are typically classified by the widths of their absorption troughs and the types of ions observed. Broad absorption lines (BALs), mini-broad absorption lines (mini-BALs), and narrow absorption lines (NALs) have trough widths of greater than 1950~km~s$^{-1}$, between 500 and 1950~km~s$^{-1}$, and less than 500~km~s$^{-1}$, respectively \cite[e.g.,][]{wey91,ald94,rei03,ves03,ham04,tru06}. If BALs from only high-ionization species (e.g., \sion{C}{iv} and \sion{N}{v}) are observed, it is designated as a HiBAL. Observing lower-ionization species such as \sion{Al}{iii} and \sion{Mg}{ii} along with high-ionization species in the outflow leads to the label LoBAL.  
%\\\\

Calculating the electron number density for an outflow typically requires observing troughs from excited and resonant state transitions from the same ion. \cite{ara13} and \cite{fin14} had success in determining $R$ from spectra within the 500-1050~\AA\ rest frame wavelength range (EUV500) since numerous excited and resonant state transitions reside in this range. The data presented here is from a spectroscopic survey of 10 quasars with known outflows (redshifts around 1) taken during Cycle 24 aimed at probing the EUV500. 

\cite{ara13} listed dozens of transitions within the EUV500 from very high-ionization potential ions that are typical of species seen in X-ray warm absorbers \cite[e.g.,][]{rey97,kaa00,cre03,kaa14}. These very high-ionization potential ions provide a link between X-ray warm absorbers and ultraviolet (UV) active galactic nucleus (AGN) outflows \cite[][]{ara13}. Many of these absorption lines have yet to be detected. In this paper, we will show clear detections for some of these previously undetected absorption lines (including \sion{Ca}{iv-v}, \sion{Ca}{v*}, and \sion{Ca}{vii*-viii*}) in addition to lines from very high-ionization potential ions.
%\\\\

This paper is part of a series of publications describing the results
of \textit{Hubble Space Telescope} (\textit{HST}) program GO-14777. 
\\
Paper I summarizes the results
for the individual objects and discusses their importance to various
aspects of quasar outflow research. 
\\
Paper II \cite[][]{xu20a} gives the full
analysis for four outflows detected in SDSS J1042+1646, including the
largest kinetic luminosity ($10^{47}$~erg~s$^{-1}$) outflow measured to date
at $R=800$~pc and another outflow at $R=15$~pc. 
\\
Paper III \cite[][]{mil20a} analyzes four outflows
detected in 2MASS J1051+1247, which show remarkable similarities, are
situated at $R\sim200$~pc, and have a combined $\dot{E}_k=10^{46}$ erg
s$^{-1}$.
\\
Paper IV \cite[][]{xu20b} presents the largest
velocity shift and acceleration measured to date in a broad absorption line (BAL) outflow.  
\\
Paper V is this work. 
\\
Paper VI \cite[][]{xu20c} analyzes two outflows
detected in SDSS J0755+2306, including one at $R=1600$~pc with
$\dot{E}_k=10^{46}-10^{47}$~erg~s$^{-1}$. 
\\
Paper VII (Miller et al. 2020c, in preparation) discusses the other
objects observed by program GO-14777, whose outflow characteristics
make the analysis more challenging.
%\\\\

The structure of this paper is as follows. Section~\ref{sec:od} presents the \textit{HST}/Cosmic Origins Spectrograph \cite[COS;][]{gre12} observations of PKS J0352-0711, which cover the diagnostic-rich wavelength range blue-ward of the Lyman limit for this quasar. Section~\ref{sec:od} also discusses the spectral fitting for the continuum and emission lines. Section~\ref{sec:ap} contrasts the amount of information contained within the EUV500 and the majority of ground-based observations with $\lambda >$~1050~\AA\ (rest frame). Section~\ref{sec:da} details the extraction of the ionic column densities, photoionization modeling, and electron number density determinations. Our results and discussions on the physical properties, distances, and energetics of each outflow are in Sections~\ref{sec:rd}~and~\ref{sec:ds}, respectively. A summary with conclusions is in Section~\ref{sec:sc}. We adopt an $h = 0.696$, $\Omega_m = 0.286$, and $\Omega_\Lambda = 0.714$ cosmology throughout this paper and use Ned Wright's Javascript Cosmology Calculator website \cite[][]{wri06}.

\section{Observations, Data Reduction, and Spectral Fitting}
\label{sec:od}
\begin{figure*}
\includegraphics[trim=2mm 2mm 1mm 3mm,clip,scale=1.0]{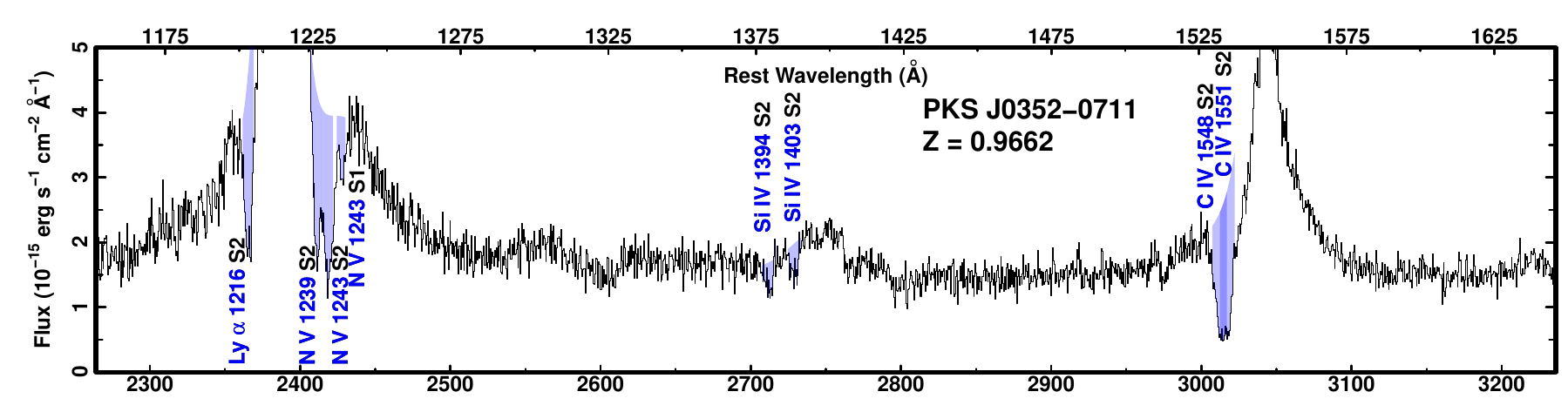}
\includegraphics[trim=2mm 2mm 1mm 3mm,clip,scale=1.0]{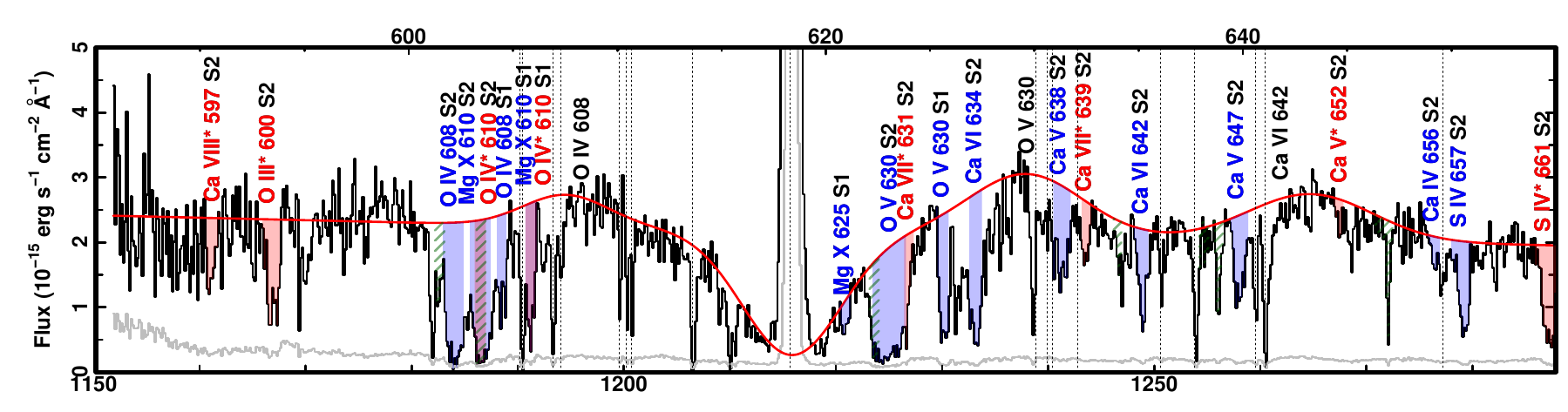}
\includegraphics[trim=2mm 2mm 1mm 3mm,clip,scale=1.0]{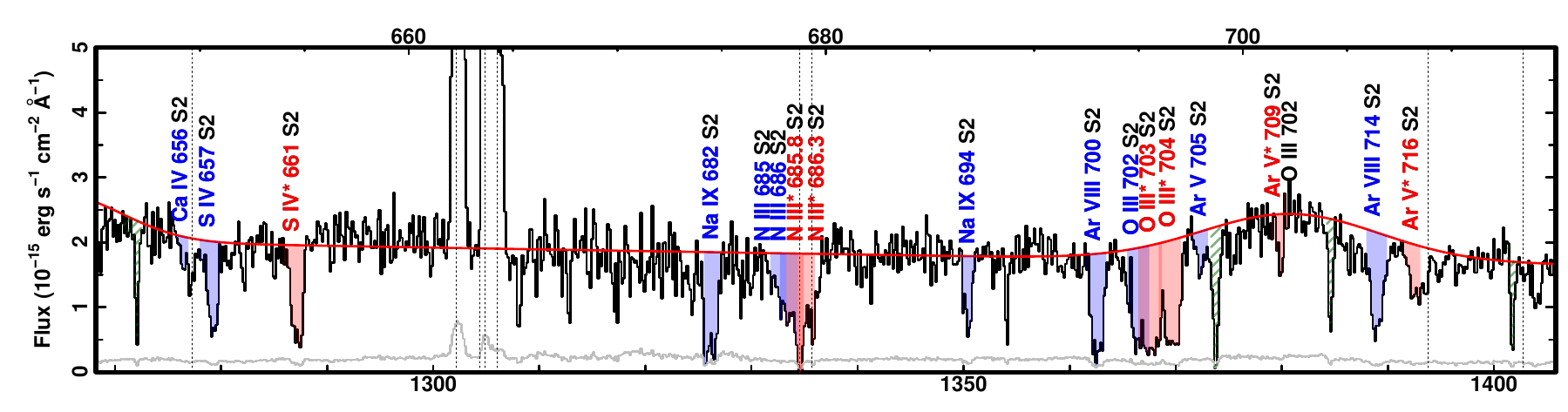}
\includegraphics[trim=2mm 2mm 1mm 3mm,clip,scale=1.0]{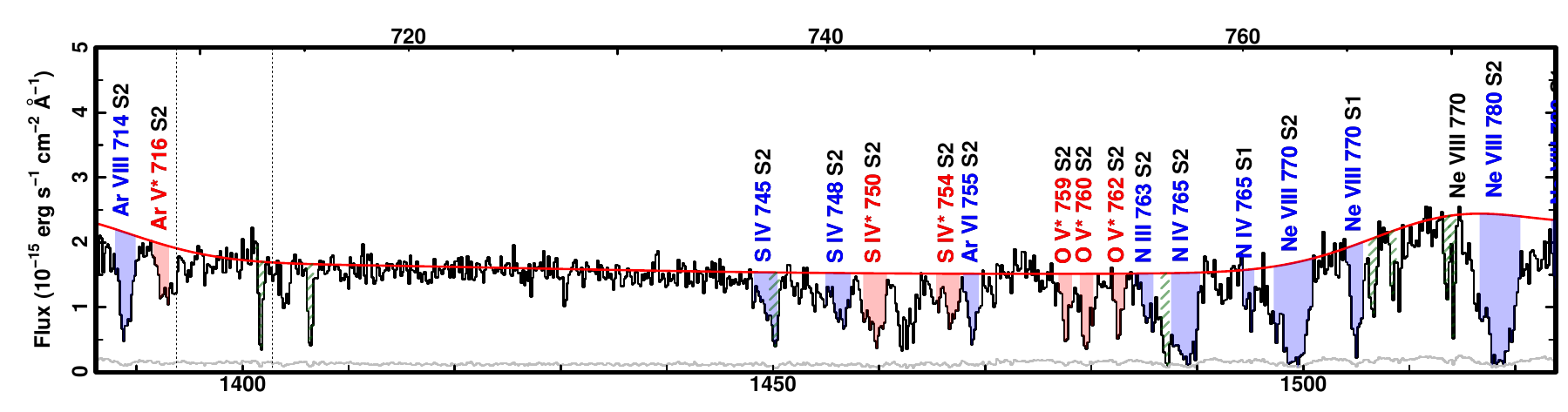}
\includegraphics[trim=2mm 5.5mm 1mm 3mm,clip,scale=1.0]{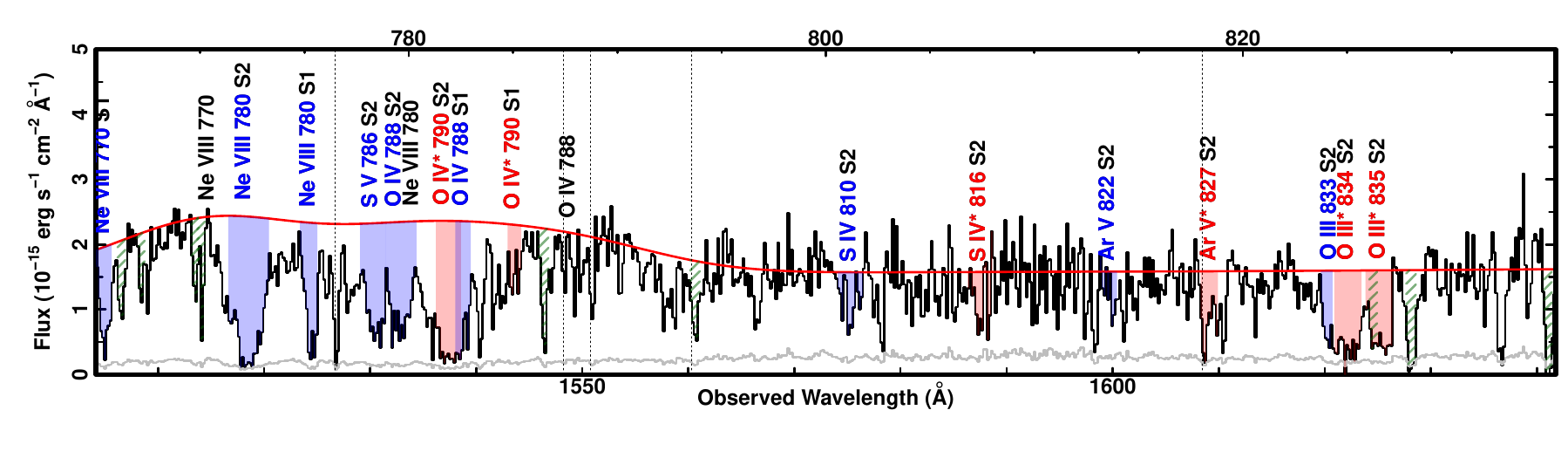}
\caption{\footnotesize{On top is a portion of the 1993 \textit{HST}/FOS spectrum in black with troughs labeled and marked with blue shaded regions from the --3150~km~s$^{-1}$ (S2) and --1950~km~s$^{-1}$ (S1) outflows. The rest are a portion of the dereddened, 2017 \textit{HST}/COS spectrum (in black) with errors (in gray). The main absorption troughs are labeled for S2 and S1. Blue shaded regions mark transitions from resonance absorption lines and red are for excited ones. Absorption troughs from intervening systems are the slanted, dark green shaded regions and the vertical dashed lines mark Galactic absorption and geocoronal emission features. The red contour traces the unabsorbed emission model. Emission lines are labeled in black.}}
\label{fig:spectrum}
\end{figure*}

PKS J0352-0711 ($z$~=~0.9662, J2000: R.A.~=~03:52:30.55, decl.~=~$-$07:11:02.3) was observed by \textit{HST}/COS in August of 2017 (PID 14777). Table~\ref{tab:obs} contains the details of each observation. The data were processed in the same way as described in \cite{mil18} and corrected for Galactic extinction with \textit{E}(\textit{B--V}) = 0.0686 \cite[]{sch11}. The bottom four panels of figure~\ref{fig:spectrum} show the dereddened, one-dimensional spectrum in black and errors in gray. Absorption troughs for the two outflow systems are delineated with S2 and S1 for $v$~=~$-$3150~km~s$^{-1}$ and $v$~=~$-$1950~km~s$^{-1}$, respectively. We use the scheme of Paper I to classify each outflow. The widest absorption troughs for each outflow are from \sion{Ne}{viii} 770.41~\AA\ and 780.32~\AA\ with widths of $\sim$500~km~s$^{-1}$ (S1) and $\sim$1400~km~s$^{-1}$ (S2), classifying both outflows as mini-BALs. S2 contains the absorption troughs that have been previously undetected: \sion{O}{III*} 599.59~\AA, \sion{Ca}{iv} 656.00~\AA, \sion{Ca}{v} 637.92~\AA\ and 646.53~\AA, \sion{Ca}{v*} 651.53~\AA, \sion{Ca}{vii*} 630.54~\AA\ and 639.15~\AA, and \sion{Ca}{viii*} 596.94~\AA. Intervening H absorption systems are also identified with slanted, dark green shaded regions.  
%\\\\

In fitting the unabsorbed emission model following the methodology of \cite{mil18}, it became apparent that the continuum emission was ill-fitted with a power law. Therefore, a cubic spline was used instead. Line emission features were modeled with Gaussian profiles. The Gaussian fits were constrained by the red side of each line since most absorption occurs on the blue side of any given emission line. Each emission line had the Gaussian centroid fixed at the rest frame wavelength. The adopted, unabsorbed emission model is shown in Figure~\ref{fig:spectrum} as a solid red contour.%For the case of the \sion{Ne}{viii} $\lambda\lambda$ 770, 780 emission doublet, the 780 line was assumed to have the same Gaussian fit as the 770 line for reasons discussed in Section ({\color{red}FILL IN}). 
\begin{deluxetable}{lcc}
\tablecaption{\textit{HST}/COS observations from 2017 August 5$^{th}$ for PKS J0352-0711.\label{tab:obs}}
\tablewidth{0pt}
\tabletypesize{\footnotesize}
\tablehead{
}
\startdata
\textit{HST}/COS grating & G130M & G160M \\
Exposure time (s) & 4072 & 4664 \\
Observed range (\AA) & 1150$-$1445 & 1400$-$1780 \\
Rest-frame range (\AA) & 585$-$735 & 710$-$905 \\
\enddata
\end{deluxetable}

\section{Contrasting the Analysis Potential of the EUV500 and $\lambda >$ 1050~\AA}
\label{sec:ap}
The top panel of figure~\ref{fig:spectrum} shows a portion of the 1993 spectra taken of PKS J0352-0711 by the Faint Object Spectrograph (FOS) aboard \textit{HST} and reported in the \textit{HST}/FOS Spectral Atlas by \cite{bec02}. This spectra covers the rest frame wavelength range of 1130--1665~\AA\ and covers much of the useful $\lambda >$ 1050~\AA\ rest frame region observed in high-ionization outflows. The majority of ground-based observations show only the absorption troughs seen in the figure for high-ionization outflows. They are usually saturated \cite[e.g.,][]{bor12b}, resulting in lower limits to their ionic column densities. Of the troughs shown for S2, the \sion{Si}{iv} troughs yield an ionic column density (\sub{N}{ion}) measurement while the rest give lower limits. Only lower bounds to the hydrogen column density (\sub{N}{H}) and \sub{U}{H} can be found with such limited data. These troughs are all from resonance transitions, making \sub{n}{e} determinations (needed to calculate $R$ and the energetics) impossible. Therefore, limited insight into the physical parameters and geometry of S2 can be obtained from such data. There is only a hint that S1 exists with the \sion{N}{V} 1242.80~\AA\ trough.
%\\\\

In contrast, the EUV500 data contains troughs from not only high-ionization potential ions (e.g., \sion{O}{iii}, \sion{O}{iv}, and \sion{N}{iii}) like those present in the FOS data but also very high-ionization potential ions (e.g., \sion{Mg}{x}, \sion{Na}{ix}, and \sion{Ne}{viii}). Such ions have ionization potentials typical of X-ray warm absorbers observed in nearby Seyfert galaxies. Several troughs will be shown to yield \sub{N}{ion} measurements, constraining \sub{U}{H} and \sub{N}{H}. There are numerous troughs from excited state transitions, yielding measurements or constraints on \sub{n}{e} for an outflow. Thus, the distance and energetics can be determined, yielding a full physical picture of the outflows as well as the potential for AGN feedback. This is made possible by the higher density of observable transitions in the EUV500, resulting in a factor of six increase in detected troughs (from 8 to about 50) over a similar $\Delta\lambda$/$\lambda$ covered by the FOS and COS observations.
%It was later observed by SDSS at the end of 2000 as part of the Legacy program. Inspection of the SDSS spectra (1940$-$3690~\AA\ rest frame) shows mini-BAL absorption for S2 from line transitions of the low ionization potential ion \sion{Mg}{ii}. However, no absorption troughs from low ionization potential ions are seen in Figure~\ref{fig:spectrum} or the \textit{HST}/FOS spectra. This suggests that the ionization state of S2 has changed between each observed epoch. Therefore, 

\section{Data Analysis}
\label{sec:da}

\subsection{Ionic Column Density Measurements}
\label{sec:cd}
The column density for a particular ion was measured by using two methods: (1) the apparent optical depth (AOD) method, and (2) the partial covering (PC) method \cite[e.g.,][]{mil18}. Therefore, a single \sub{N}{ion} for a given ionic energy state is determined by one ionic transition (AOD method) or two ionic transitions (PC method). The PC method yields reliable results when multiple lines from the same ionic energy state and different oscillator strengths have different trough depths, allowing for the measurement of a viable PC solution. Table~\ref{tab:col} lists the total (sum of all observed ionic energy states) column density for each ion in both outflow systems. The ratio of the measured column densities to the predicted column densities from the best-fit model are also given (see Section~\ref{sec:photmod} and Figure~\ref{fig:sol}). When the measured \sub{N}{ion} are lower limits, we expect this ratio to be less than one and vice versa for upper limits. Many excited states have multiple transitions with small wavelength separations ($<$~0.5~\AA). In such cases, we combine each set of transitions into a single transition for labeling in Figure~\ref{fig:spectrum}. A list of atomic data for the transitions can be found in Table 3 of Paper II. 
%\\\\

Non-black saturation is a concern, so we use the scheme used in Paper II to decide on ionic column density measurements, lower limits, and upper limits: PC \sub{N}{ion} are measurements, regions where the maximum optical depth, $\tau_{max}$, is less than 0.05 are taken as upper limits; troughs that have 0.05 $< \tau_{max} <$ 0.5 with other troughs from similar ionization potential ions that have $\tau_{max} >$ 2 are also taken as measurements; and all others are lower limits. We report both the upper and lower limit \sub{N}{ion} for \sion{S}{iv} of S2. They are measured from the \sion{S}{iv} 810 and \sion{S}{iv*} 816 regions (upper limit) and the \sion{S}{iv} 657 and \sion{S}{iv*} 661 troughs (lower limit). The \sion{S}{iv} troughs around 750~\AA\ are blended with unknown absorption, yielding unreliable PC \sub{N}{ion}. Following previous works \cite[e.g.,][]{mil18,xu18}, the adopted value is chosen to be the PC \sub{N}{ion} when available, or else it is the AOD \sub{N}{ion} limits. To account for systematics in the unabsorbed emission model, all adopted error values (see Table~\ref{tab:col}) have added an additional 20\% error in quadrature \cite[e.g.,][]{mil18,xu18}. 
%From equation (9) in \cite{sav91}, the AOD ionic column density can be calculated from
%\begin{equation}\label{eq:sav91}
%\sub{N}{ion}=\frac{m_e c}{\pi e^2 f \lambda} \int \tau(v)dv
%\end{equation} where $m_e$ is the mass of the electron, $c$ is the speed of light, $e$ is the electric charge, $\sub{N}{ion}$ is the ionic column density, $f$ is the oscillator strength, $\lambda$ is the wavelength, and $\tau(v)$ is the velocity dependent optical depth. Similarly from \cite{ara99}, the PC ionic column density can be determined from 
%\begin{equation}
%\sub{N}{ion}=\frac{m_e c}{\pi e^2 f \lambda} \int C(v)\tau(v)dv
%\end{equation}
%where $C(v)$ is the velocity dependent effective covering factor. All line properties are from the National Institute of Standards and Technology (NIST) online database except the oscillator strengths for Ca and Ar of which are from \cite{fis06} and \cite{ver96}. 

\begin{deluxetable}{ccccc}
\tablecaption{Total Ionic Column Densities\label{tab:col}}
\tablewidth{0pt}
\tabletypesize{\footnotesize}
\tablehead{
\colhead{Ion} & \colhead{AOD} & \colhead{PC} & \colhead{Adopted} & \colhead{\scriptsize$\frac{\textnormal{Adopted}}{\textnormal{Best Model}}$} \\ 
 & \colhead{($10^{12} $cm$^{-2}$)} & \colhead{($10^{12} $cm$^{-2}$)} & \colhead{($10^{12} $cm$^{-2}$)} & 
}
\startdata
\multicolumn{5}{c}{v = --3150~km~s$^{-1}$}\\
\tableline
\sion{N}{iii} & 890\pme{170}{150} & \nodata & \color{blue}\textgreater890\me{230} & \textgreater0.39\me{0.17}\\
\sion{N}{iv} & 610\pme{50}{30} & \nodata & \color{blue}\textgreater610\me{130} & \textgreater0.02\me{0.01}\\
\sion{O}{iii} & 7700\pme{450}{260} & \nodata & \color{blue}\textgreater7700\me{1600} & \textgreater1.90\me{0.74}\\
\sion{O}{iv} & 13000\pme{810}{500} & \nodata & \color{blue}\textgreater13000\me{2700} & \textgreater0.10\me{0.04}\\
\sion{O}{v} & 4600\pme{590}{260} & \nodata & \color{blue}\textgreater4600\me{940} & \textgreater0.02\me{0.01}\\
\sion{Ne}{viii} & 15000\pme{850}{460} & \nodata & \color{blue}\textgreater15000\me{3000} & \textgreater0.07\me{0.03}\\
\sion{Na}{ix} & 2000\pme{210}{190} & 2700\pme{960}{340} & 2700\pme{1100}{650} & 2.11\pme{2.28}{0.86}\\
\sion{S}{iii} & 60\pme{10}{10} & \nodata & \color{red}\textless60\pe{20} & \textless0.30\pe{0.31}\\
\sion{S}{iv} & 340\pme{20}{10} & \nodata & \color{blue}\textgreater340\me{70} & \textgreater0.11\me{0.04}\\
\sion{S}{iv} & 1600\pme{250}{170} & \nodata & \color{red}\textless1600\pe{410} & \textless0.50\pe{0.52}\\
\sion{S}{v} & 160\pme{10}{10} & \nodata & \color{blue}\textgreater160\me{30} & \textgreater0.03\me{0.01}\\
\sion{Cl}{v} & 100\pme{30}{20} & \nodata & \color{red}\textless100\pe{40} & \textless0.28\pe{0.30}\\
\sion{Cl}{vii} & 220\pme{30}{20} & \nodata & \color{red}\textless220\pe{50} & \textless0.47\pe{0.48}\\
\sion{K}{v} & 210\pme{70}{70} & \nodata & \color{red}\textless210\pe{80} & \textless1.96\pe{2.15}\\
\sion{K}{vi} & 1000\pme{230}{170} & \nodata & \color{red}\textless1000\pe{310} & \textless2.79\pe{2.89}\\
\sion{K}{vii} & 250\pme{70}{70} & \nodata & \color{red}\textless250\pe{90} & \textless0.35\pe{0.37}\\
\sion{K}{ix} & 200\pme{40}{30} & \nodata & \color{red}\textless200\pe{60} & \textless0.59\pe{0.62}\\
\sion{Ar}{iv} & 880\pme{260}{210} & \nodata & \color{red}\textless880\pe{310} & \textless1.46\pe{1.56}\\
\sion{Ar}{v} & 1100\pme{140}{110} & \nodata & \color{red}\textless1100\pe{270} & \textless0.28\pe{0.29}\\
\sion{Ar}{vi} & 1500\pme{170}{70} & \nodata & \color{blue}\textgreater1500\me{290} & \textgreater0.30\me{0.12}\\
\sion{Ar}{viii} & 930\pme{50}{40} & 1400\pme{290}{140} & 1400\pme{390}{330} & 0.15\pme{0.16}{0.06}\\
\sion{Ca}{iv} & 1100\pme{290}{190} & \nodata & \color{blue}\textgreater1100\me{280} & \textgreater2.62\me{1.15}\\
\sion{Ca}{v} & 9100\pme{550}{590} & \nodata & \color{blue}\textgreater9100\me{1900} & \textgreater4.52\me{1.78}\\
\sion{Ca}{vi} & 2500\pme{170}{230} & \nodata & \color{blue}\textgreater2500\me{670} & \textgreater0.39\me{0.16}\\
\sion{Ca}{vii} & 1500\pme{120}{180} & \nodata & \color{blue}\textgreater1500\me{370} & \textgreater0.06\me{0.02}\\
\sion{Ca}{viii} & 1400\pme{280}{180} & \nodata & \color{blue}\textgreater1400\me{330} & \textgreater0.06\me{0.03}\\
\sion{Mg}{x} & 6700\pme{500}{350} & \nodata & \color{blue}\textgreater6900\me{1400} & \textgreater0.41\me{0.16}\\
\tableline
\multicolumn{5}{c}{v = --1950~km~s$^{-1}$}\\
\tableline
\sion{N}{iii} & 80\pme{30}{30} & \nodata & \color{red}\textless80\pe{40} & \textless4.53\pe{4.60}\\
\sion{N}{iv} & 130\pme{30}{20} & \nodata & \color{blue}\textgreater130\me{30} & \textgreater0.25\me{0.12}\\
\sion{O}{iv} & 1500\pme{260}{150} & 1700\pme{290}{190} & 1700\pme{450}{380} & 1.08\pme{1.12}{0.43}\\
\sion{O}{v} & 410\pme{20}{20} & \nodata & \color{blue}\textgreater410\me{160} & \textgreater0.05\me{0.02}\\
\sion{Ne}{viii} & 4000\pme{290}{190} & \nodata & \color{blue}\textgreater4000\me{810} & \textgreater0.35\me{0.14}\\
\sion{Na}{ix} & 140\pme{70}{70} & \nodata & \color{red}\textless140\pe{80} & \textless1.28\pe{1.28}\\
\sion{Mg}{x} & 1500\pme{220}{260} & 1700\pme{100}{190} & 1700\pme{330}{400} & 0.95\pme{0.97}{0.38}\\
\sion{S}{iv} & 7\pme{2}{3} & \nodata & \color{red}\textless7\pe{2} & \textless0.73\pe{0.77}\\
\sion{Ar}{vi} & 90\pme{50}{50} & \nodata & \color{red}\textless90\pe{60} & \textless0.83\pe{1.01}\\
\sion{Ar}{viii} & 80\pme{20}{20} & \nodata & \color{red}\textless80\pe{30} & \textless1.07\pe{1.20}\\
\sion{Ca}{vii} & 320\pme{120}{140} & \nodata & \color{red}\textless320\pe{130} & \textless0.99\pe{1.08}\\
\enddata
\tablecomments{Total ionic column densities (excited plus resonance, where applicable) for each outflow system with the measured and adopted values and errors. Adopted values in blue are lower limits, upper limits are in red, and measurements are in black. The ratio of the adopted values to the column densities from the best-fit Cloudy model are in the last column and those errors also account for the uncertainty in metallicity of each element.}
%\tablenotetext{a}{The ratio of the adopted value with the best model value determined by the photoionization solution.}
\end{deluxetable}

\subsection{Photoionization Modeling}
\label{sec:photmod}
Given that the troughs of each outflow are narrow and any blended troughs do not hinder our analysis, we follow the methodology of previous works \cite[e.g.,][]{mil18,xu18,xu19} and not the SSS method in Paper II. To determine the \sub{N}{H} and \sub{U}{H} that best model the outflow system, a grid of Cloudy \cite[][version c17.00]{fer17} photoionization models were generated. We used two metallicities and three spectral energy distributions (SEDs): the UV-soft SED \cite[]{dun10}, the HE0238 SED \cite[]{ara13}, and the MF87 SED \cite[]{mat87}. These three SEDs were chosen since they give a representative range of SED shapes that are commonly attributed to radio quiet quasars \cite[]{ara13}.
%\\\\

The two metallicities are solar \cite[Z$_{\astrosun}$,][]{gre10} and one super-solar (Z = 4.68 Z$_{\astrosun}$, see Table~\ref{tab:met}). These were chosen since chemical abundances of outflow systems have been shown to be between solar and 4-5 times solar \cite[e.g.,][Arav et al. 2020b, in preparation]{gab06,ara07}. The super-solar abundances for C, N, O, Mg, Si, Ca, and Fe are from Table 2 of \cite{bal08} for the 10$^{11}$ M$_{\astrosun}$ bulge mass. We note that their quoted metallicity (7.22 Z$_{\astrosun}$) is likely in error as we calculate 4.1 Z$_{\astrosun}$ assuming all other elements remain solar and 4.68 with the enhancement of all other elements described next. The abundances of the other elements were chosen to be increased above solar by a factor similar to the elements from \cite{bal08} that come from the same fusion sources \cite[i.e., Ne-Al and P have sources of C and Ne while Si and S-Ca have sources of O and Si; e.g.,][]{arn96,and89} instead of a simple linear increase with Z \cite[e.g.,][]{ham93}. Since we are uncertain of both the super-solar abundances from \cite{bal08} given the metallicity discrepancy and our values for the other elements, we assume an abundance uncertainty in each element of 50\%. This uncertainty is based on the relative error between the metallicity given by \cite{bal08} and what we determine, i.e. (7.22--4.68)/4.68 $\approx$ 50\%.
%\\\\

\begin{figure}
	\includegraphics[scale=0.305]{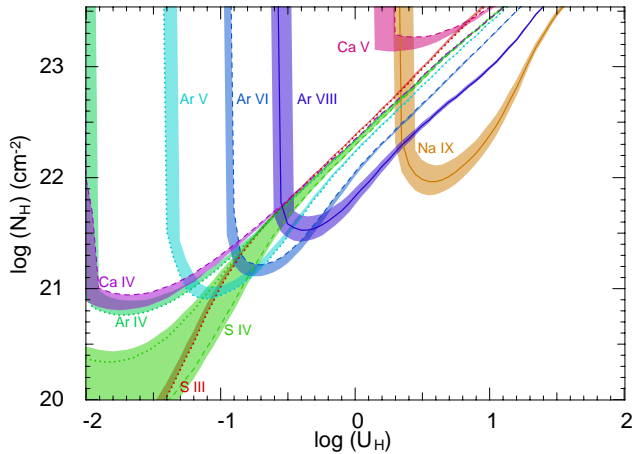}
	\caption{\footnotesize{Colored contours show the model parameters that are consistent with the observed values assuming the HE0238 SED and solar metallicity. Solid contours represent ionic column densities taken as measurements, and dotted or dashed contours represent \sub{N}{ion} upper or lower limits, respectively. The shaded bands are the 1$\sigma$ uncertainties for each contour (see Table~\ref{tab:col}). For clarity's sake, only a subset of all ions are shown. Any solution that matches the lower limit column density of \sion{Ca}{v} overpredicts the column densities of \sion{Ar}{iv}, \sion{Ar}{v}, \sion{Ar}{viii}, and \sion{S}{iv} by up to a factor of 10. Invoking a super-solar metallicity is a possible solution.}}
	\label{fig:nosol}
\end{figure}

For a particular pair of \sub{N}{H} and \sub{U}{H}, ionic column densities from the model are compared to the measured counterparts. In Figure~\ref{fig:nosol}, the colored contours for individual ions show where each model-predicted \sub{N}{ion}, assuming the HE0238 SED and solar metallicity, is consistent ($<$ 1$\sigma$) with the corresponding observed values for S2. The colored contours with solid lines are ionic column densities treated as measurements and dotted or dashed lines are \sub{N}{ion} upper or lower limits, respectively. It is evident that for solar metallicity, there is no viable solution. Any solution that matches the lower limit column density of \sion{Ca}{v} will simultaneously overpredict the column densities of \sion{Ar}{iv}, \sion{Ar}{v}, \sion{Ar}{viii}, and \sion{S}{iv} by up to two orders of magnitude. The same is true for the other two SEDs. One possible solution is to invoke a super-solar metallicity to reduce the \sub{N}{H} required to match the observations of \sion{Ca}{v}. 
%\\\\

\begin{figure}
\includegraphics[scale=0.305]{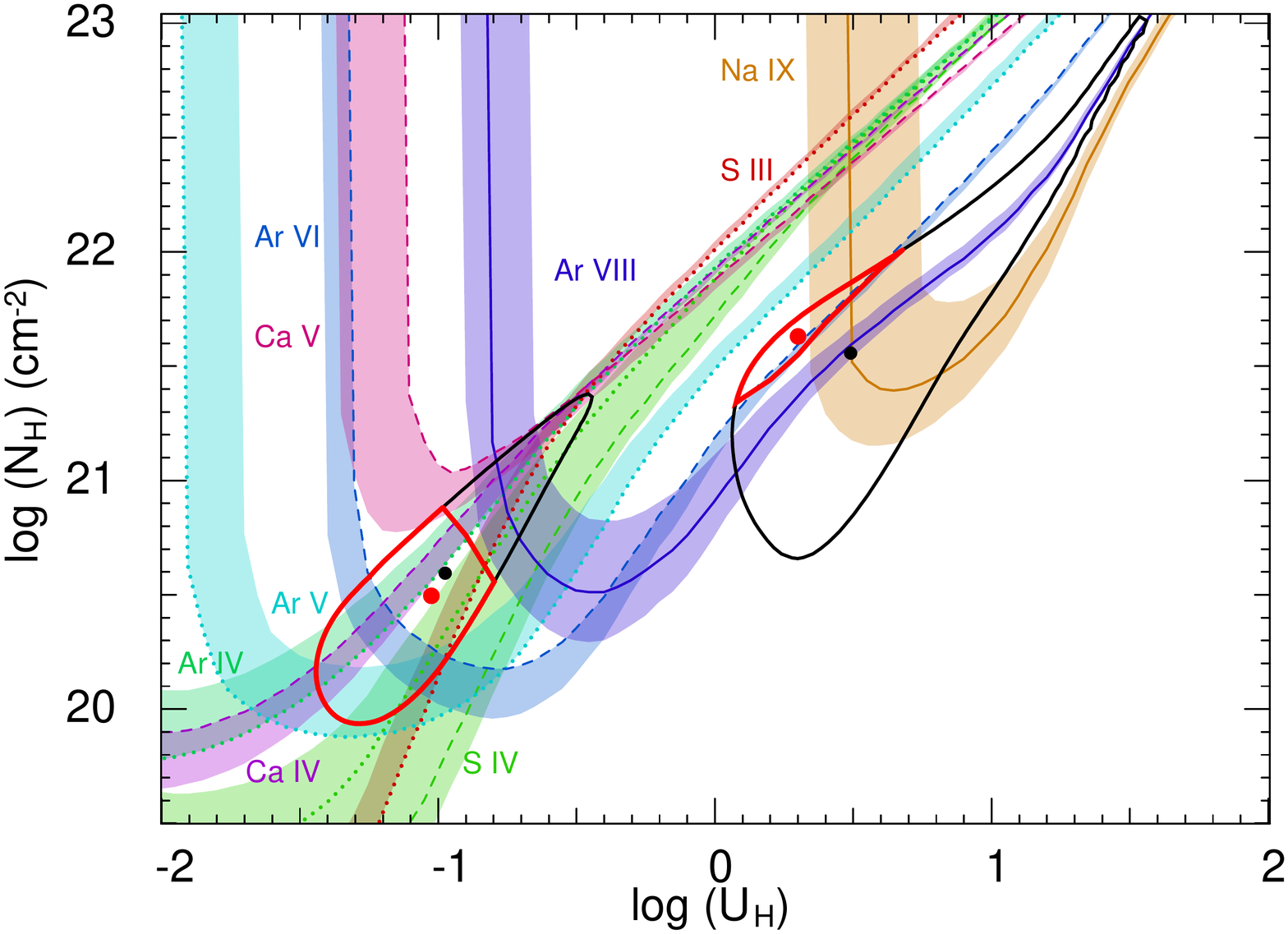}\\
\includegraphics[scale=0.305]{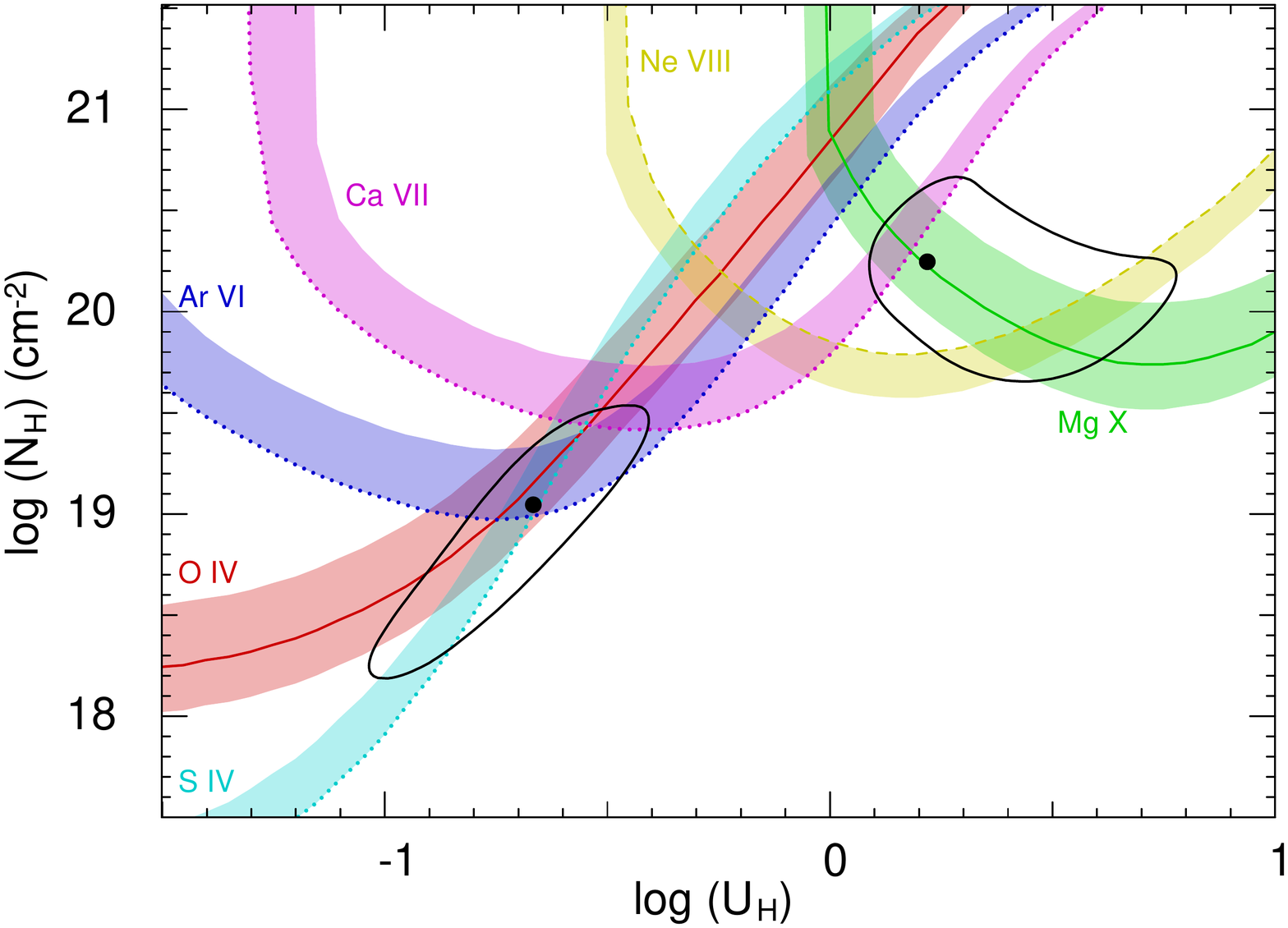}
\caption{\footnotesize{Colored contours show the model parameters that are consistent with the observed values assuming the HE0238 SED and Z = 4.68 Z$_{\astrosun}$ (see Table~\ref{tab:met}). For clarity's sake, only a subset of all ions are shown. Solid contours represent \sub{N}{ion} measurements, while dotted and dashed contours are \sub{N}{ion} upper and lower limits, respectively. The shaded bands are the 1$\sigma$ uncertainties for each contour, including the uncertainty in the metallicity (see Tables~\ref{tab:col}~and~\ref{tab:met}). \textit{Top:} Two-phase photoionization solution for the S2 outflow system. The black dots are the $\chi^2$-minimization solutions for each ionization phase based on the total ionic column densities, and the ellipses encircling them are their 1$\sigma$ uncertainties. The red dots and corresponding ellipses take into account the \sion{O}{v*} column density constraint and are chosen as the best solution (see Section~\ref{sec:photmod}). \textit{Bottom:} Two-phase photoionization solution for the S1 outflow system.  The black dots are the best $\chi^2$-minimization solutions for each ionization phase, and the ellipses encircling them are their 1$\sigma$ uncertainties.}}
\label{fig:sol}
\end{figure}

Figure~\ref{fig:sol} shows the same contours for both S1 and S2, assuming the HE0238 SED but with Z = 4.68 Z$_{\astrosun}$. The best-fit solution is determined through $\chi^2$-minimization of the model-predicted \sub{N}{ion} compared to the measured ionic column densities (all values in Table~\ref{tab:col} when accounting for the uncertainty in the metallicity). The solutions and corresponding 1$\sigma$ uncertainties are the black dots and ellipses. However, for S2, we also take into account that the \sion{O}{V*} column density primarily comes from the very high-ionization phase (see Section \ref{sec:veloff}), which shifts the solution to the red dots and constrains the errors to the red ellipses (the overlap of the 1$\sigma$ contours for \sion{O}{v*} and the black ellipses). Imposing the same \sion{O}{v*} constraint and using the other SEDs yields no overlap between the 1$\sigma$ contours for \sion{O}{v*} and the black ellipses, i.e., a worse solution for the very high-ionization phase. Therefore, the adopted best-fit solutions for S1 and S2 are those with the HE0238 SED and super-solar metallicity since we assume both outflows have the same incident SED and metallicity.
%\\\\
\begin{deluxetable}{ccc}
\tablecaption{Z = 4.68 Z$_{\astrosun}$ Composition\label{tab:met}}
\tablewidth{0pt}
\tabletypesize{\footnotesize}
\tablehead{
\colhead{Element} & \colhead{\textit{X}/\textit{H}} & \colhead{\textit{X}/\textit{X}$_{\astrosun}$}
}
\startdata
\color{red}C	&	(5.4$\pm$2.70) x 10$^{-4}$	& 2.00$\pm$1.00\\
\color{red}N	&	(5.0$\pm$2.50) x 10$^{-4}$	& 7.41$\pm$3.71\\
\color{red}O	&	(1.5$\pm$0.75) x 10$^{-3}$	& 3.02$\pm$1.51\\
Ne				&	(3.4$\pm$1.70) x 10$^{-4}$	& 4.00$\pm$2.00\\
Na				&	(6.9$\pm$3.45) x 10$^{-6}$	& 4.00$\pm$2.00\\
\color{red}Mg	&	(2.0$\pm$1.00) x 10$^{-4}$	& 4.90$\pm$2.45\\
Al				&	(1.2$\pm$0.60) x 10$^{-5}$	& 4.00$\pm$2.00\\
\color{red}Si	&	(4.4$\pm$2.20) x 10$^{-4}$	& 13.49$\pm$6.75\\
P				&	(1.0$\pm$0.50) x 10$^{-6}$	& 4.00$\pm$2.00\\
S				&	(1.3$\pm$0.65) x 10$^{-4}$	& 10.00$\pm$5.00\\
Cl				&	(3.2$\pm$1.60) x 10$^{-6}$	& 10.00$\pm$5.00\\
Ar				&	(2.5$\pm$1.25) x 10$^{-5}$	& 10.00$\pm$5.00\\
K				&	(1.1$\pm$0.55) x 10$^{-6}$	& 10.00$\pm$5.00\\
\color{red}Ca	&	(2.5$\pm$1.25) x 10$^{-5}$	& 11.48$\pm$5.74\\
\color{red}Fe	&	(3.7$\pm$1.85) x 10$^{-4}$	& 11.74$\pm$5.87\\
\enddata
%\tablenotetext{a}{}
\tablecomments{Red elements are from \citet[][]{bal08}.}
\end{deluxetable}
Both outflow systems require a two-phase photoionization solution to satisfy the column densities from both the very high-ionization potential ions and high-ionization potential ions \cite[see][]{ara13}. A single phase solution at the intersection of the \sion{O}{iv} and \sion{Mg}{x} contours for S1 overpredicts the upper limit column density of \sion{Ar}{vi} by over a factor of five. Similarly, the very high-phase solution of S2 produces negligible amounts of the \sion{S}{iv} column density, and the high-phase solution fails to reproduce the observed column density of \sion{Na}{ix}, necessitating a two-phase solution. However, even this two-phase solution under predicts the column density of \sion{Ca}{v} by almost a factor of five ($< 2\sigma$) and overpredicts the column density of \sion{S}{iii} by a factor of three ($\simeq 2\sigma$) and the column density of \sion{Ar}{viii} by nearly a factor of seven ($\simeq 5\sigma$). However, the discrepancy with \sion{Ar}{viii} depends on our estimate for the abundance of \sion{Ar}{viii}, of which an enhancement of only twice solar would reduce the difference to within 2$\sigma$. 

\subsection{Determining \sub{n}{e}}
\label{sec:ed}
All of the excited state troughs shown in Figure~\ref{fig:spectrum} become populated through collisions with free electrons. These collisions depend on both the electron number density and gas temperature. Therefore, calculating the relative populations between an excited and resonance or two excited states will yield \sub{n}{e} \cite[e.g.,][]{dek01,ham01,dek02,kor08}. Following the methodology of previous works \cite[e.g.,][]{bor12b,ara13,ara15,ara18,cha15b}, we used the CHIANTI 8.0.7 database \cite[]{der97,lan13} to calculate the predicted population ratios from the states of each ion. This ratio is equal to the ratio of the measured \sub{N}{ion}. 
%\\\\

However, not all of the observed excited states are useful for this approach. \sion{Ca}{vii*} 630.54~\AA\ is blended with \sion{O}{v} 629.73~\AA. \sion{N}{iii*} 685.82~\AA\ and 686.34~\AA\ are heavily blended with not only each other and \sion{N}{iii} 685.52~\AA, but also with Galactic absorption from \sion{C}{ii} 1334.53~\AA\ and \sion{C}{ii*} 1335.71~\AA. Comparing troughs from the same energy level for a particular ion, \sion{Ca}{V} 637.92~\AA\ and 646.53~\AA; \sion{O}{iii} 702.34~\AA\ and 832.93~\AA; \sion{O}{iv} 608.40~\AA\ and 787.71~\AA; \sion{O}{iii*} 702.90~\AA, 703.85~\AA, 833.75~\AA, and 835.29~\AA; and \sion{O}{iv*} 609.83~\AA\ and 790.20~\AA\ of the S2 outflow system exhibit 1:1 trough depths. The PC method is unusable in these instances since the troughs are saturated and the PC method needs at least one trough from a given ionic energy state to be shallower. This leaves \sion{O}{V*} 759.44~\AA, 760.45~\AA, and 762.00~\AA; \sion{S}{iv*} 661.40~\AA, 750.22~\AA, and 753.76~\AA\ (with \sion{S}{iv} 657.32~\AA, 744.90~\AA, and 748.39~\AA); \sion{Ca}{viii*} 596.94~\AA; and \sion{Ca}{vii*} 639.15~\AA\ as potentially useful density diagnostics for S2 and \sion{O}{iv*} 790.20~\AA\ (with \sion{O}{iv} 608.40~\AA\ and 787.71~\AA) for S1.
%\\\\

\begin{figure}
	\includegraphics[scale=0.33,angle=-90]{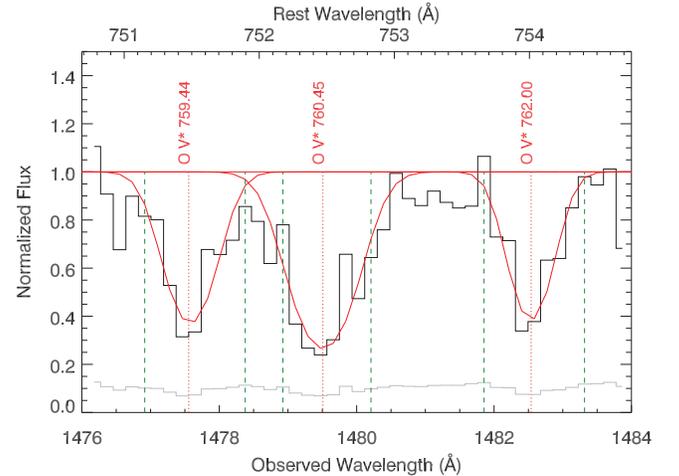}%
	\caption{\footnotesize{Best-fitting Gaussian profiles for the \sion{O}{V*} 760.45 \AA\ ($J=2$), 762.00 \AA\ ($J=2$) and 759.44 \AA\ ($J=0$) absorption troughs in red are overlaid on the data in black. The vertical red dotted lines mark the velocity centroid, and the vertical green dashed lines show the fitting range.}}
	\label{fig:fit}
\end{figure}

\begin{figure}
	\includegraphics[scale=0.33,angle=-90]{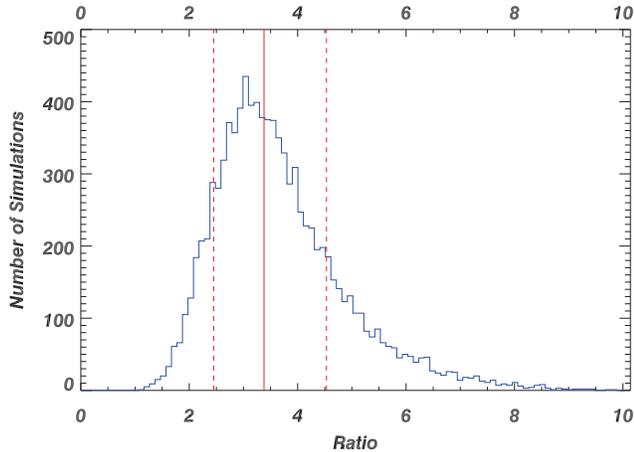}%
	\caption{\footnotesize{Monte Carlo ratio distribution for the \sion{O}{V*} 760.45 \AA\ ($J=2$), 762.00 \AA\ ($J=2$) and 759.4 \AA\ ($J=0$) absorption troughs. The solid red line is the ratio determined from the best-fitting Gaussian profiles, and the dashed red lines mark the 1$\sigma$ boundaries of the distribution.}}
	\label{fig:mc}
\end{figure}

\begin{figure}
	\includegraphics[trim=4mm 7mm 10mm 5mm,clip,scale=0.34]{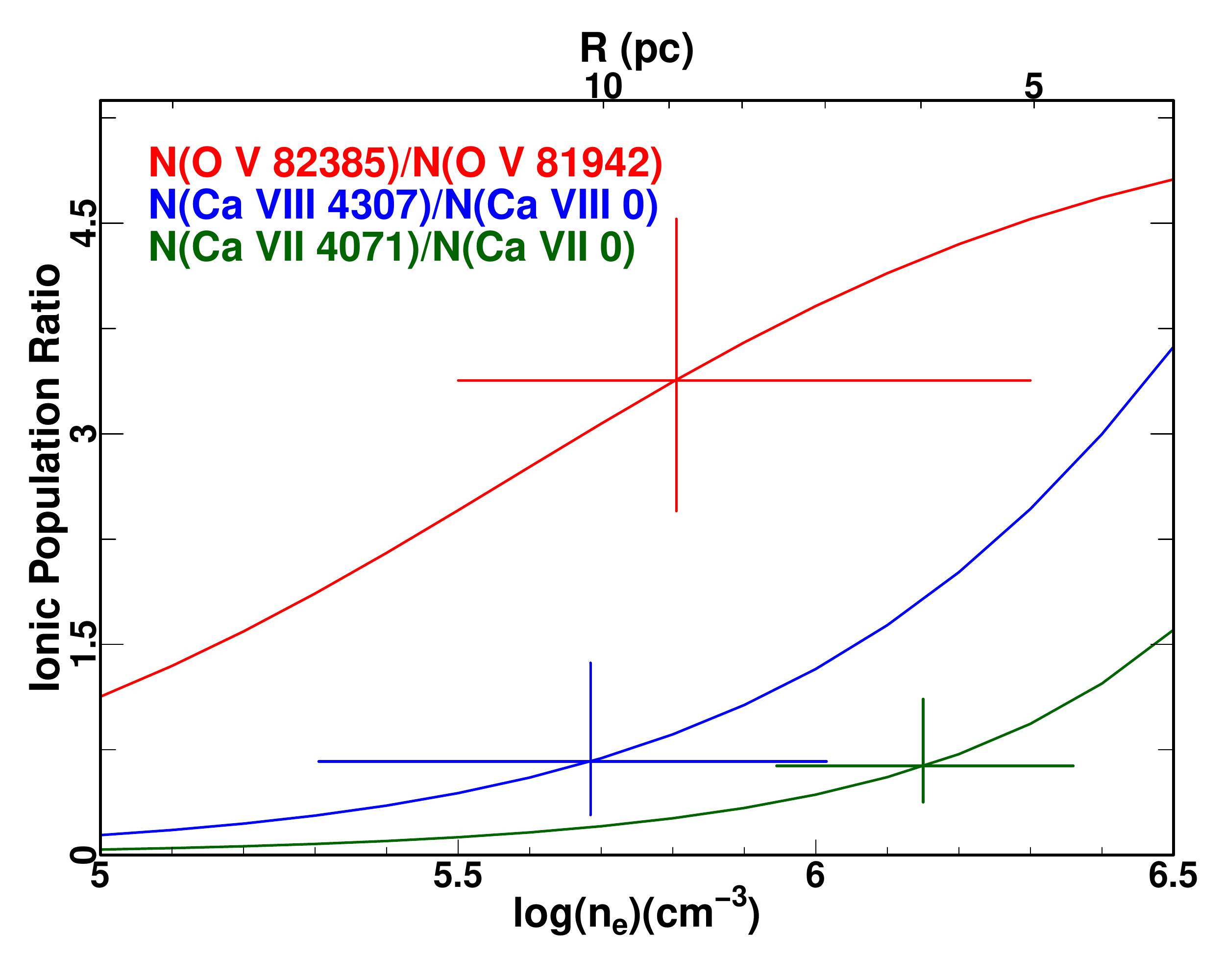}\\
	\includegraphics[trim=4mm 7mm 10mm 5mm,clip,scale=0.34]{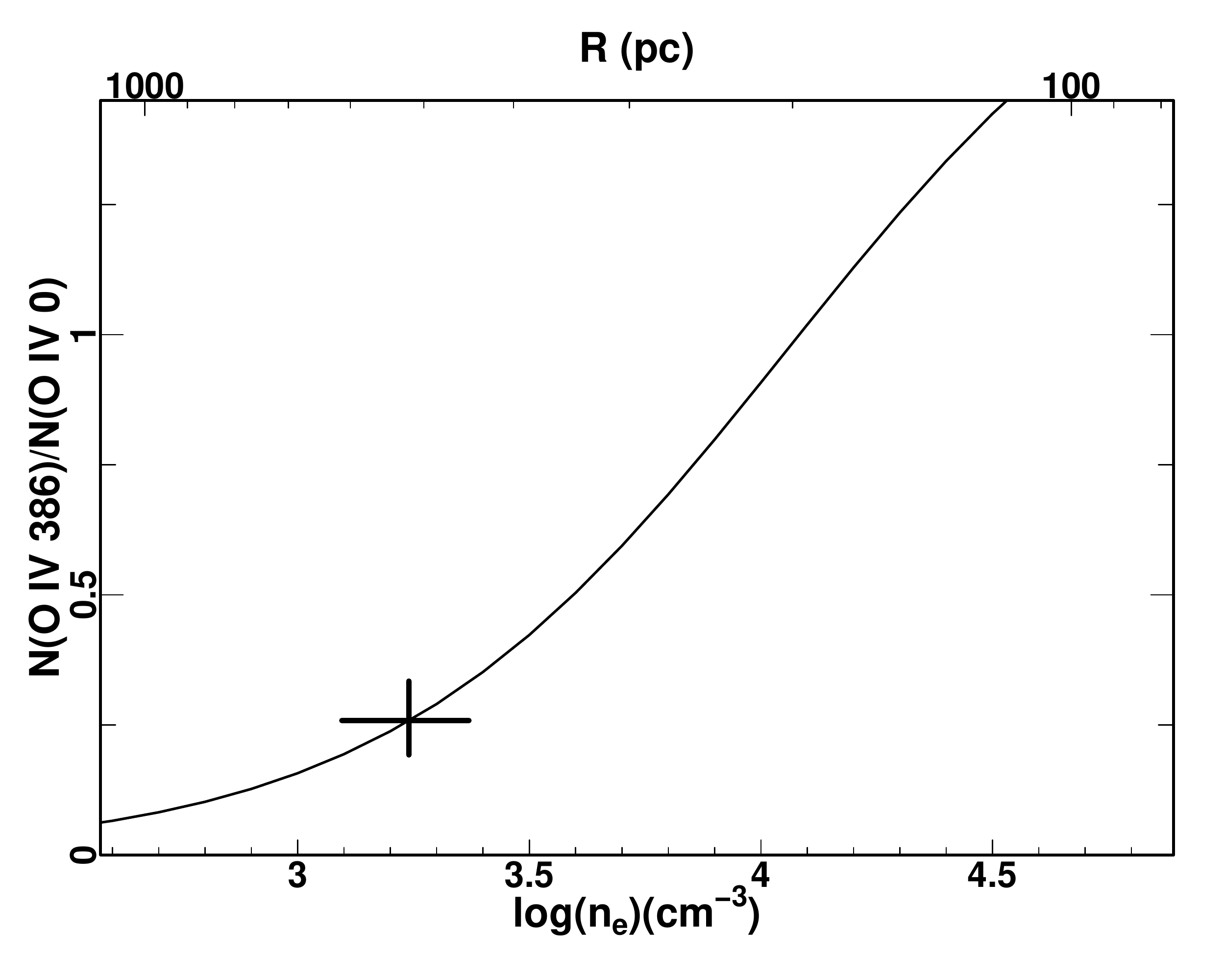}
	\caption{\footnotesize{\textit{Top:} the electron number density, \sub{n}{e}, and the distance from the central source, $R$ (equation (\ref{eq:R})), of the S2 outflow system based on the lines of \sion{O}{v}, \sion{Ca}{vii}, and \sion{Ca}{viii}. The average temperature from the photoionization solution for the very high-ionization phase is 52,700~K. The ratios and CHIANTI contours of \sion{Ca}{vii} and \sion{Ca}{viii} have been scaled up by a factor of 10 for clarity's sake. The distance on the top axis assumes the \sion{O}{v*} \sub{n}{e} and \sub{U}{H} of the very high-phase solution. \textit{Bottom:} \sub{n}{e} for the S1 outflow system based on the \sion{O}{iv} ratio. The assumed temperature is 15,900~K. The distance axis assumes the \sub{U}{H} of the high-phase solution. See section~\ref{sec:ed}.}}
	\label{fig:dens}
\end{figure}

For \sion{O}{V*}, the line at 759.44~\AA\ is from the $J=0$ (81,942 cm$^{-1}$) energy level and the other two lines are from the $J=2$ (82,385 cm$^{-1}$) energy level (for additional lines and transition parameters, see Paper II). Therefore, using the $J=2$ lines with the PC method yields one column density, and assuming the same covering solution, the column density for the $J=0$ energy level and subsequent ratio can be calculated. We assume the temperature of the very high-ionization phase (52,700~K) for reasons discussed in Section~\ref{sec:veloff}. However, given the signal-to-noise ratio of the data, directly measuring the ionic column density from the data yielded a ratio with large enough errors that only a lower limit on $n_e$ could be determined, i.e., the ratio is consistent with the theoretical Boltzmann limit ($\simeq$ 5 in the top panel of Figure~\ref{fig:dens}). To obtain a better constrained \sub{n}{e}, we fit each of the absorption troughs using Gaussian optical depth profiles over the velocity range $-3300$ to $-3000$~km~s$^{-1}$ (See Figure~\ref{fig:fit}): 
\begin{equation}
\tau_i(v) = \frac{A_i}{\sigma_i\sqrt{2\pi}}*exp\bigg(\frac{(v-v_i)^2}{2\sigma_i^2}\bigg)
\end{equation}
\begin{equation}
I_i(v) = exp(-\tau_i)
\end{equation}
where for trough i, $A_i$ is scaling factor, $\sigma_i$ is the velocity dispersion (FWHM = $2\sqrt{2ln(2)}\sigma$), $v_i$ is the velocity centroid, and $I_i(v)$ are the fitted, normalized flux values. The fitting parameters for each trough were allowed to vary independent of each other, resulting in nine parameters with associated errors (See Table~\ref{tab:fit}). The same PC procedure outlined above can then be used on these fitted functions to get the ratio. To propagate errors, we used a Monte Carlo approach, randomly choosing each parameter from a normal distribution 10,000 times and calculating the final ratio. The distribution of the ratios is shown in Figure~\ref{fig:mc}, and we adopt $\frac{N(J=2)}{N(J=0)} =$  3.4\pme{1.1}{0.9}.
\begin{deluxetable}{cccc}
\tablecaption{Best-fitting Gaussian Parameters for the \sion{O}{v*} Multiplets.\label{tab:fit}}
\tablewidth{0pt}
\tabletypesize{\footnotesize}
\tablehead{
	\colhead{Line} & \colhead{$A_i$} & \colhead{$\sigma_i$} & \colhead{$v_i$} \\
	\colhead{} & \colhead{(km s$^{-1}$)} & \colhead{(km s$^{-1}$)} & \colhead{(km s$^{-1}$)}
}
\startdata
%\sion{O}{v*} 759.4	&	170.4$\pm$22.7	& 68.3$\pm$9.7	& -3158.4$\pm$10.5\\
%\sion{O}{v*} 760.4	&	279.7$\pm$40.2	& 84.1$\pm$15.2	& -3161.7$\pm$13.8\\
%\sion{O}{v*} 762	&	140.6$\pm$18.1	& 58.5$\pm$8.0	& -3163.7$\pm$8.7\\
\sion{O}{v*} 759.4	&	170$\pm$23	& 68$\pm$10	& -3158$\pm$11\\
\sion{O}{v*} 760.4	&	280$\pm$40	& 84$\pm$15	& -3162$\pm$14\\
\sion{O}{v*} 762	&	141$\pm$18	& 59$\pm$8	& -3164$\pm$9\\
\enddata
%\tablecomments{The assumed composition for Z = 4.67Z$_{\astrosun}$.}
\end{deluxetable}
%\\\\

In the top panel of Figure~\ref{fig:dens}, the red contour shows the expected column density ratio as a function of electron number density for \sion{O}{v*} as determined by CHIANTI for a temperature of 52,700~K, which is determined by the Cloudy solution for the very high-phase. Overlaid on that contour is the measured column density ratio and uncertainties from the Monte Carlo results. We calculate an \sion{O}{v*} derived \sub{n}{e} for the very high-ionization phase of S2 to be log(\sub{n}{e}) = 5.8\pme{0.5}{0.3}~cm$^{-3}$. Since the velocity centroids of the troughs from the high- and very high-ionization phase solutions of S2 are similar, it is very likely that they are at the same distance. For the two phases to be located at the same distance, the high-phase must have an $n_e$ larger by a factor equal to the ratio of the two ionization parameters (see equation~\ref{eq:R}): log(\sub{n}{e}) = 7.1\pme{0.8}{0.4}~cm$^{-3}$. The \sion{S}{iv*} and \sion{S}{iv} absorption troughs are primarily produced by the high-ionization phase and yield a log(\sub{n}{e}) $>$ 5, which is consistent with our assumed value of 7.1.
%\\\\

There are two additional \sub{n}{e} diagnostics for S2 that can be calculated from \sion{Ca}{vii*} and \sion{Ca}{viii*}. We first used the model-predicted values for the total \sub{N}{ion} of \sion{Ca}{vii} and \sion{Ca}{viii} along with our measured \sub{N}{ion} of \sion{Ca}{vii*} and \sion{Ca}{viii*} to estimate the ground state populations of each ion. This is possible since the very high-ionization phase produces over 90\% of both total \sub{N}{ion}. Then we calculated the ratio of each excited state to the estimated ground state. Plotting these with their CHIANTI contours as seen in Figure~\ref{fig:dens} shows consistent values for \sub{n}{e} between all three diagnostics.
%\\\\

\begin{figure}
	\includegraphics[trim=4mm 7mm 10mm 19mm,clip,scale=0.34]{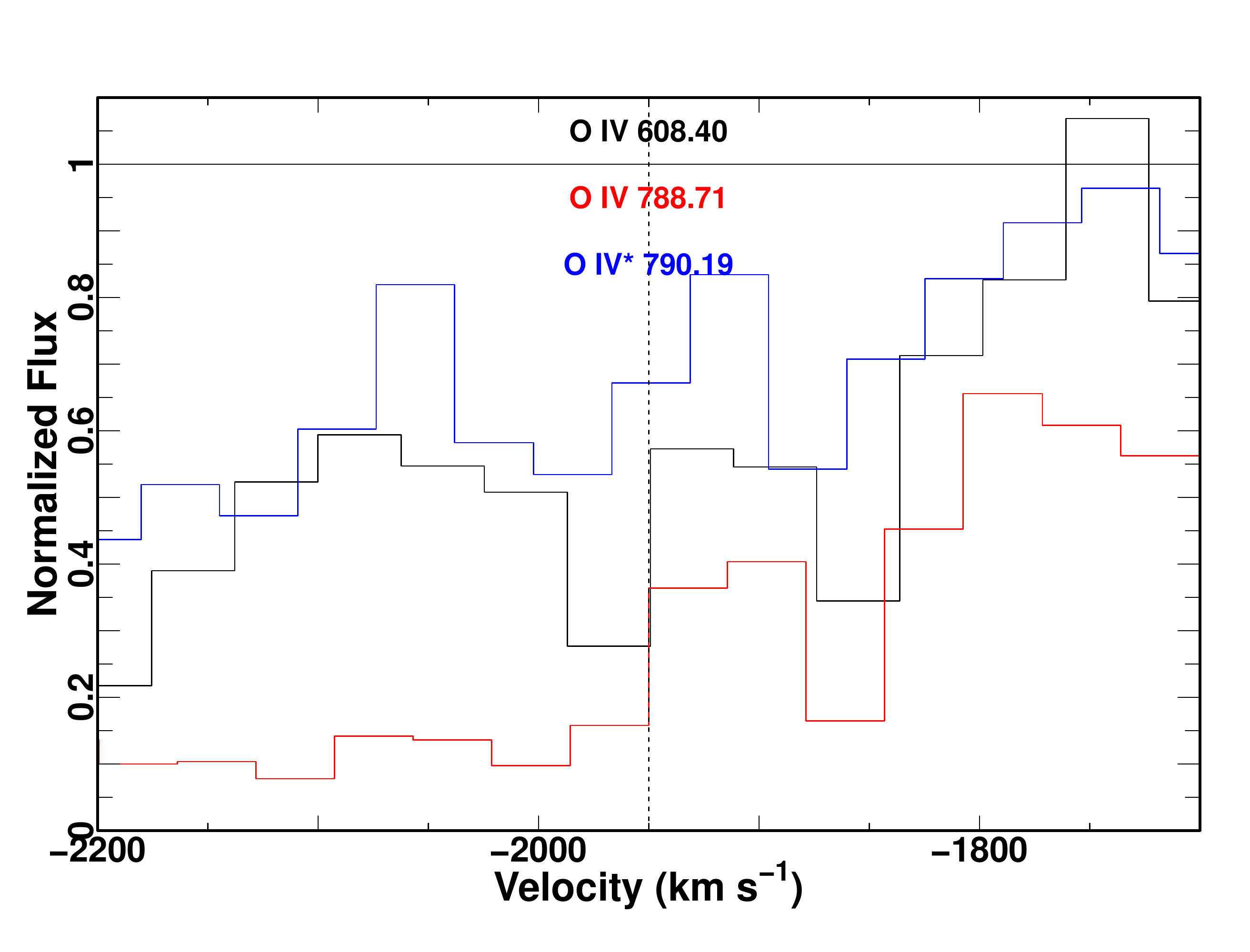}
	\caption{\footnotesize{Overlap of the \sion{O}{iv} 608.40~\AA, \sion{O}{iv} 788.71~\AA, and \sion{O}{iv*} 790.19~\AA\ troughs for S1. To the left of the vertical, dotted line (-1950~km~s$^{-1}$) is the blend of the \sion{O}{iv*} 790.20~\AA\ trough of S2 with the \sion{O}{iv} 788.71~\AA\ trough of S1. To the right of the line is not blended, allowing for the ionic column density of the resonance lines to be determined.}}
	\label{fig:oiv}
\end{figure}

For the \sion{O}{iv} and \sion{O}{iv*} troughs of S1, the \sion{O}{iv} 608.40~\AA\ and \sion{O}{iv*} 790.19~\AA\ lines are not blended with any other lines, but the \sion{O}{iv} 787.71~\AA\ line is blended on the blue side with the \sion{O}{iv*} 790.19~\AA\ line of S2 (See Figure~\ref{fig:oiv}). Since the \sion{O}{iv*} trough is shallower than the \sion{O}{iv} 787.71~\AA\ trough and they have the same oscillator strength value, the \sub{N}{ion} of \sion{O}{iv*} is less than the \sub{N}{ion} of \sion{O}{iv}. Therefore, \sub{n}{e} for S1 is smaller than the critical density of log(\sub{n}{e,crit}) = 4.1 for this diagnostic \cite[see][]{ara18}. We assume the trough is symmetric and double the red half PC \sub{N}{ion} value for the \sion{O}{iv} \sub{N}{ion}. We use the covering solution of the \sion{O}{iv} 608.40~\AA\ and 787.71~\AA\ doublet along with the \sion{O}{iv*} 790.19~\AA\ trough to determine the \sion{O}{iv*} ionic column density in the same way as above since the \sion{O}{iv*} 609.83~\AA\ line is blended with the \sion{Mg}{x} 609.79~\AA\ line. \sion{O}{iv} is produced by the high-phase, of which has a gas temperature of 15,900~K as determined by the Cloudy model solution. The bottom panel of Figure~\ref{fig:dens} shows the resulting ratio, and we calculate log(\sub{n}{e}) = 3.2\pme{0.2}{0.1}~cm$^{-3}$ for the high-phase. There are no density diagnostic troughs for the very high-phase of S1.

%% In this section, we use  the \subsection command to set off
%% a subsection.  \footnote is used to insert a footnote to the text.

%% Observe the use of the LaTeX \label
%% command after the \subsection to give a symbolic KEY to the
%% subsection for cross-referencing in a \ref command.
%% You can use LaTeX's \ref and \label commands to keep track of
%% cross-references to sections, equations, tables, and figures.
%% That way, if you change the order of any elements, LaTeX will
%% automatically renumber them.

%% This section also includes several of the displayed math environments
%% mentioned in the Author Guide.

\section{Results}
\label{sec:rd}

\subsection{Outflow Properties, Distance, and Energetics}
From the definition of the ionization parameter, we can determine the distance each outflow is from the central source: 
\begin{equation}
\label{eq:R}
\sub{U}{H} = \frac{\sub{Q}{H}}{4\pi R^2\ssub{n}{H} c}
\end{equation}
where $R$ is the distance from the central source, \ssub{n}{H}~is the hydrogen number density ($\sub{n}{e} \approxeq 1.2\ssub{n}{H}$ for highly ionized plasma), $c$ is the speed of light, and $\sub{Q}{H}$ is the ionizing hydrogen photon rate. $\sub{Q}{H}$ was calculated by integrating the HE0238 SED for energies above 1 Ryd, yielding $\sub{Q}{H}=3.1\times 10^{56}$~s$^{-1}$. Under the assumption of a partially filled, thin shell outflow presented by \cite{bor12}, the mass flow rate and kinetic luminosity are given by, respectively,
\begin{equation}
\label{eq:M}
\dot{M}\simeq 4\pi \Omega R N_H \mu m_p v 
\end{equation}
and
\begin{equation}
\label{eq:E}
\dot{\sub{E}{K}}\simeq \frac{1}{2} \dot{M} v^2
\end{equation}
where $R$ is the distance from the central source, $\mu = 1.4$ is the mean atomic mass per proton, \sub{N}{H} is the hydrogen column density, $m_p$ is the proton mass, $v$ is the outflow velocity, and $\Omega$ is the global covering factor. Since we can not directly measure how much the outflow covers the source, we use the frequency of \sion{Ne}{viii} mini-BAL outflow detections as a proxy, i.e., $\Omega$ = 0.40\pme{0.14}{0.14} \cite[][]{muz13}. Table~\ref{tab:res} contains the physical properties, energetics, and distances for each outflow system. As can be seen, S2 is fairly close to the central source at 8.9~pc while S1 is much farther out at 520~pc. 
\begin{deluxetable}{lcccc}
\tabletypesize{\footnotesize}
\tablecaption{Physical Properties, Distances, and Energetics of the two outflow systems\label{tab:res}}
\tablewidth{0pt}
\tablehead{
\colhead{Outflow System} & \multicolumn{2}{c}{$-$3150~km~s$^{-1}$ (S2)} & \multicolumn{2}{c}{$-$1950~km~s$^{-1}$ (S1)}\\
\colhead{Ionization Phase} & \colhead{Very High} & \colhead{High} & \colhead{Very High} & \colhead{High}
}
\startdata
log$_{}$(\sub{N}{H}) & 21.63$^{+0.27}_{-0.30}$ & 20.50$^{+0.38}_{-0.66}$ & 20.25$^{+0.42}_{-0.59}$ & 19.05$^{+0.49}_{-0.86}$ \\
\ [cm$^{-2}$]\\
\tableline
log$_{}$(\sub{U}{H}) & 0.3$^{+0.4}_{-0.2}$ & -1.0$^{+0.2}_{-0.5}$ & 0.2$^{+0.6}_{-0.1}$ & -0.7$^{+0.3}_{-0.3}$ \\
\ [dex]\\
\tableline
log(\sub{n}{e}) & 5.8$^{+0.5}_{-0.3}$ & \tablenotemark{a}7.1$^{+0.8}_{-0.4}$ & \tablenotemark{a}2.4$^{+0.3}_{-0.7}$ & 3.2$^{+0.2}_{-0.1}$ \\
\ [cm$^{-3}$]\\
\tableline
Distance & \multicolumn{2}{c}{8.9$^{+4.9}_{-4.5}$} & \multicolumn{2}{c}{520$^{+300}_{-150}$} \\
\ [pc]\\
\tableline
$\dot{M}$ & \multicolumn{2}{c}{7.5$^{+9.0}_{-4.9}$} & \multicolumn{2}{c}{11.2$^{+21.8}_{-8.2}$} \\
\ [$M_{\astrosun}$yr$^{-1}$]\\
\tableline
log($\dot{\sub{E}{K}}$)\tablenotemark{b} & \multicolumn{2}{c}{43.37$^{+0.34}_{-0.46}$} & \multicolumn{2}{c}{43.13$^{+0.47}_{-0.57}$}\\
\ [erg s$^{-1}$]\\
\tableline
$\dot{\sub{E}{K}}/L_{edd}$ & \multicolumn{2}{c}{0.02\pme{0.06}{0.01}} & \multicolumn{2}{c}{0.01\pme{0.04}{0.01}}\\
\ [\%]\\
\tableline
log(\ssub{f}{V}) & \multicolumn{2}{c}{-2.4\pme{0.6}{0.9}} & \multicolumn{2}{c}{-2.1\pme{0.8}{1.2}} \\
\enddata
%log$_{}$(\sub{U}{H}) & 0.30$^{+0.38}_{-0.23}$ & -1.03$^{+0.23}_{-0.47}$ & 0.22$^{+0.56}_{-0.13}$ & %-0.67$^{+0.26}_{-0.37}$ \\
%\ [dex]\\
%\tableline
%log(\sub{n}{e}) & 5.81$^{+0.50}_{-0.31}$ & \tablenotemark{a}7.13$^{+0.78}_{-0.45}$ & %\tablenotemark{a}2.35$^{+0.32}_{-0.69}$ & 3.24$^{+0.13}_{-0.15}$ \\
%\ [cm$^{-3}$]\\
%\tableline
%\tablecomments{The physical properties, distances, and energetics of the two ionization phases in each outflow system. }
\tablenotetext{a}{Assuming that both ionization components are at the same distance.}
\tablenotetext{b}{Assuming $\Omega$ = 0.37 and where $N_H$ is the sum of the two ionization phases.}
\tablecomments{Bolometric luminosity, \sub{L}{bol} = 5.5$^{+0.1}_{-0.1}\times$10$^{46}$~erg~s$^{-1}$ assuming the HE0238 SED.}
\end{deluxetable}

\subsection{Velocity Centroid Offset between Troughs in S2}
\label{sec:veloff}
\begin{figure}
	\includegraphics[trim=4mm 12mm 10mm 19mm,clip,scale=0.35]{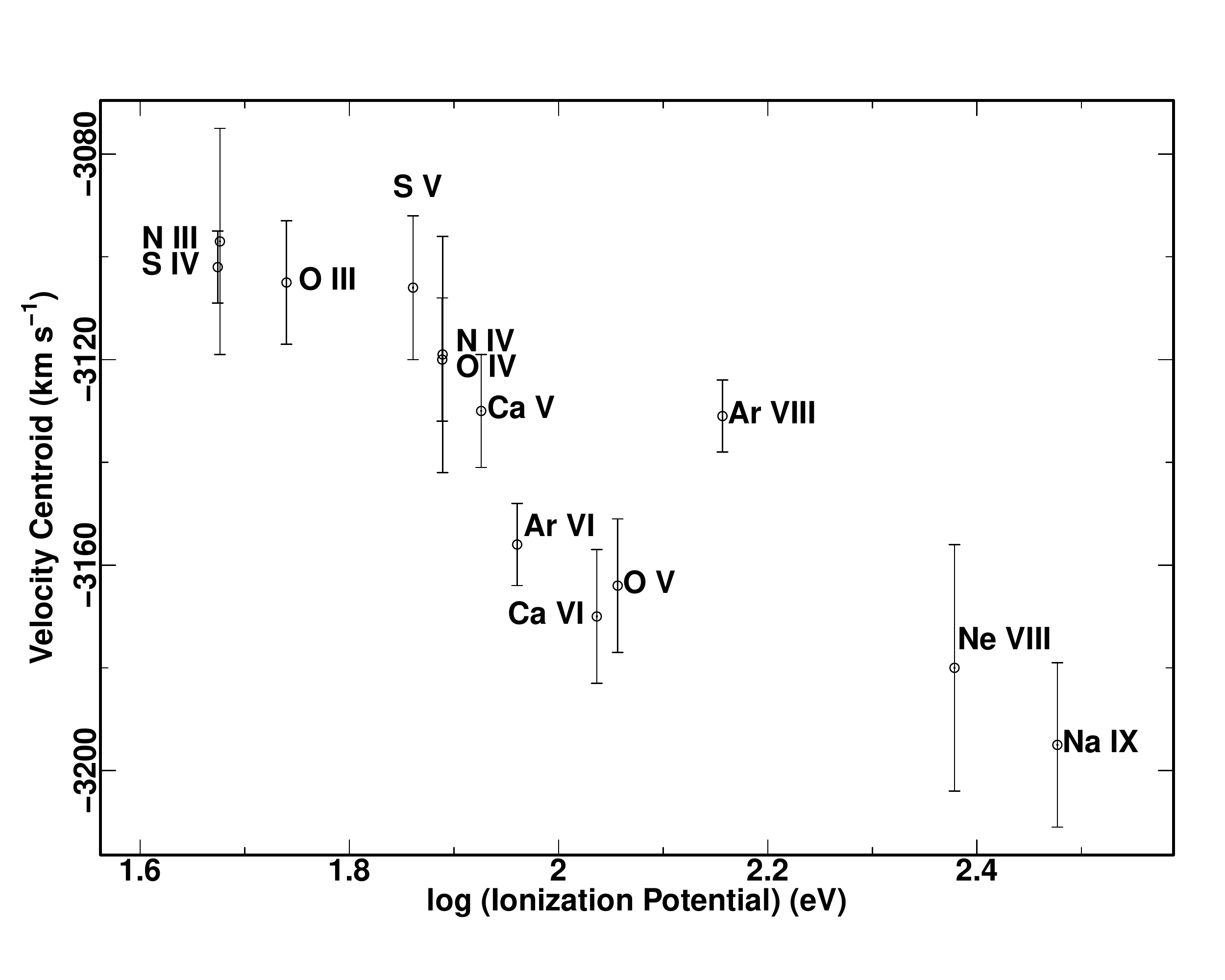}
	\caption{\footnotesize{Velocity centroid (from Gaussian profile fits) for troughs in S2 as a function of the ionization potential of each ion. There is an increase in the magnitude of the velocity for higher ionization potential ions, suggesting the two photoionization solutions are offset in velocity (see Section~\ref{sec:veloff} and Figure~\ref{fig:sol}). }}
	\label{fig:vel}
\end{figure}

In analyzing the S2 absorption troughs, we noticed a velocity centroid shift between ions of different ionization potentials. To quantify this shift, we used Gaussian optical depth profiles to fit troughs where the deepest parts had normalized flux values less than 0.5. We only fitted the sections of each trough that were not heavily contaminated due to blending or intervening systems. We also restricted each trough for a particular ion to have the same velocity centroid and simultaneously fit those troughs, minimizing the effects of spurious data or unidentified contaminations. Figure~\ref{fig:vel} shows that the fitted velocity centroid for each ion tends to increase in magnitude with increasing ionization potential. This suggests that the two photoionization solutions are offset in velocity as the column densities of \sion{Na}{ix} and \sion{S}{iv}, for example, are produced entirely by the very high-phase and high-phase, respectively. Therefore, the \sion{O}{v*} troughs, with their higher velocity centroids (see Table~\ref{tab:fit}), should primarily be produced by the very high-phase.

\section{Discussion}
\label{sec:ds}

\subsection{S2 Photoionization Solution and \sub{n}{e} Accuracy}
As shown in Section~\ref{sec:ed}, the \sub{n}{e} derived from the \sion{O}{v*}, \sion{Ca}{vii*}, and \sion{Ca}{viii*} diagnostic ratios for the very high-phase of S2 all gave consistent results. The photoionization solution for this phase, which yielded our estimates for the \sub{N}{ion} of \sion{Ca}{vii} and \sion{Ca}{viii}, was determined primarily by the constraints imposed by the \sub{N}{ion} measurements of \sion{O}{v*}, \sion{Na}{ix}, and \sion{Ar}{viii}, and not by the \sion{Ca}{vii*} and \sion{Ca}{viii*} \sub{N}{ion} lower limits. Therefore, the fact that the photoionization solution gave \sion{Ca}{vii} and \sion{Ca}{viii} \sub{N}{ion} values such that the \sub{n}{e} derived were all within errors speaks to the accuracy and robustness of the photoionization solution and \sub{n}{e}. 

\subsection{AGN Feedback Contribution}
To judge the potential for AGN feedback, kinetic luminosities exceeding 0.5\% \cite[]{hop10} or 5\% \cite[]{sca04} of the Eddington luminosity are thought to be sufficient. Using the \sion{Mg}{ii}-based equation from \cite{bah19} and their methodology for measuring the \sion{Mg}{II} emission line FWHM and local continuum level from Sloan Digital Sky Survey (SDSS) data, we estimate the mass of the super massive black hole to be 8.7\pme{11.8}{5.6}$\times 10^8 M_{\astrosun}$ (including systematics). This corresponds to an Eddington luminosity (\sub{L}{edd}) of $1.1^{+1.5}_{-0.7}\times10^{47}$~erg~s$^{-1}$. Taking the ratio of the kinetic luminosities with respect to \sub{L}{edd} gives 0.02\% and 0.01\% for S2 and S1, respectively (see Table~\ref{tab:res}). Therefore, these outflows in PKS J0352-0711 are not significant contributors to AGN feedback processes.

\subsection{Volume Filling Factor and S2 Velocity Shift}
From equation 6 in Paper II, the volume filling factor for two phases is given by 
\begin{equation}
	\ssub{f}{V} = \frac{\sub{U}{H,\tiny\textit{HP}}}{\sub{U}{H,\tiny\textit{VHP}}}\times\frac{\sub{N}{H,\tiny\textit{HP}}}{\sub{N}{H,\tiny\textit{VHP}}}
\end{equation}
where HP and VHP denote the high-phase and very high-phase, respectively (see also Section 2.5 in Paper I). S1 and S2 have \ssub{f}{V} values of 8$\times$10$^{-3}$ and 4$\times$10$^{-3}$, respectively (see Table~\ref{tab:res}). These values are similar to those in Papers II, III, IV, and VI but 3-4 orders of magnitude larger than what is seen in HE0238-1904 \cite[][]{ara13}. %where it mimics the situation seen in X-ray warm absorbers observed in Seyfert galaxies \cite[e.g.,][]{net03}. 
%\\\\

Given the ionization potential-dependent velocity shift between the troughs in S2, it warrants a closer look at the properties of S2. The very high-phase has a thickness of $\Delta R$ = \sub{N}{H}/\sub{n}{e} = 2$\times$10$^{-3}$~pc while the high-phase has $\Delta R$ = 8$\times$10$^{-6}$pc. The densities are also among the highest that have been measured to date. Therefore, it is possible that these high densities and thicknesses are related to the observed velocity shift. Paper II also shows an outflow (in SDSS J1042+1646 at $-$7500 km s$^{-1}$) with the same very high-phase \sub{n}{e} as S2 and an \sub{n}{e} for the high-phase that is half that of the high-phase \sub{n}{e} for S2. Both phases in the SDSS J1042+1646 outflow also have larger thicknesses by about five times that of the corresponding phases of S2. However, a velocity shift analysis could not be done for that outflow since the only observed high-ionization potential ions were \sion{O}{iv} and \sion{N}{iv}, and the troughs were blended and wide, making velocity centroid measurements unreliable. 
\subsection{X-Ray Warm Absorber Connection}
X-ray warm absorbers have been shown to span up to 5 orders of magnitude in the ionization parameter, i.e., -1 $<$ log($\xi$) $<$4 (for the HE0238 SED, log($\xi$) $\approxeq$ log(\sub{U}{H})+1.3), and a continuous \sub{N}{H} as a function of $\xi$ is often invoked \cite[e.g.,][]{ste03,cos07,hol07,mck07,beh09}. The necessity of the two phases to sufficiently explain the observed absorption troughs from the high and very high-ionization potential ions in PKS J0352-0711 is similar to what is seen for X-ray warm absorbers, and we can not rule out phases at higher \sub{U}{H} and \sub{N}{H} with our data. The \sub{U}{H} and \sub{N}{H} are also comparable to those determined for X-ray warm absorbers. Current X-ray observatories (\textit{XMM-Newton} and \textit{Chandra}) do not have the sensitivity to obtain useful data on outflows in luminous quasars like PKS J0352-0711. However, the future observatory Athena \cite[][]{bar17} is designed to have over 50 times the effective area (at 1 keV) for spectroscopy compared to current observatories, enabling more distant quasars to be probed. 
\subsection{The S2 Outflow ``Shading Effect" on the S1 Outflow}
\label{sec:tp}
\begin{figure*}
	\includegraphics[trim=0mm 5mm 10mm 10mm,clip,scale=0.37]{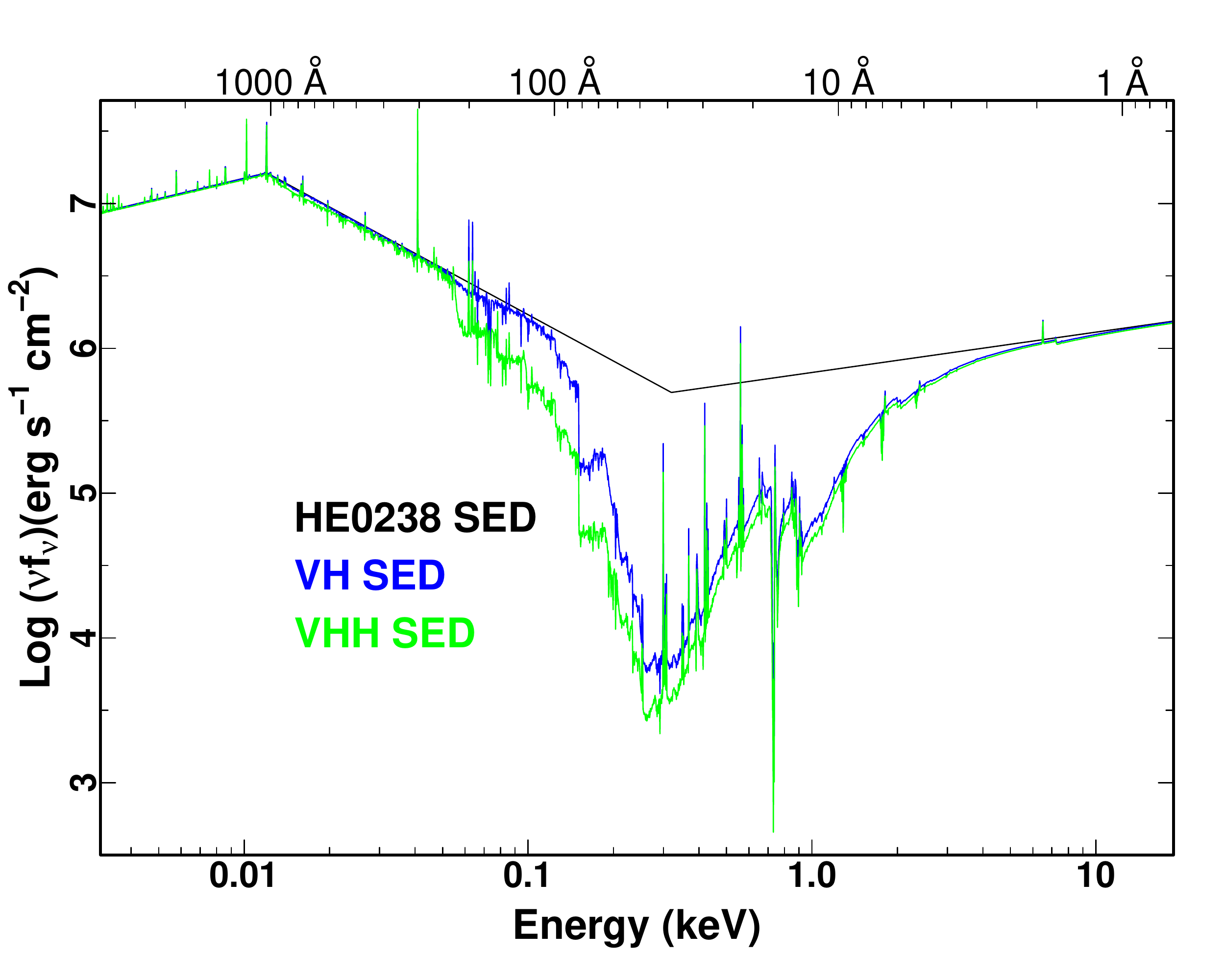}\includegraphics[trim=0mm 5mm 10mm 10mm,clip,scale=0.37]{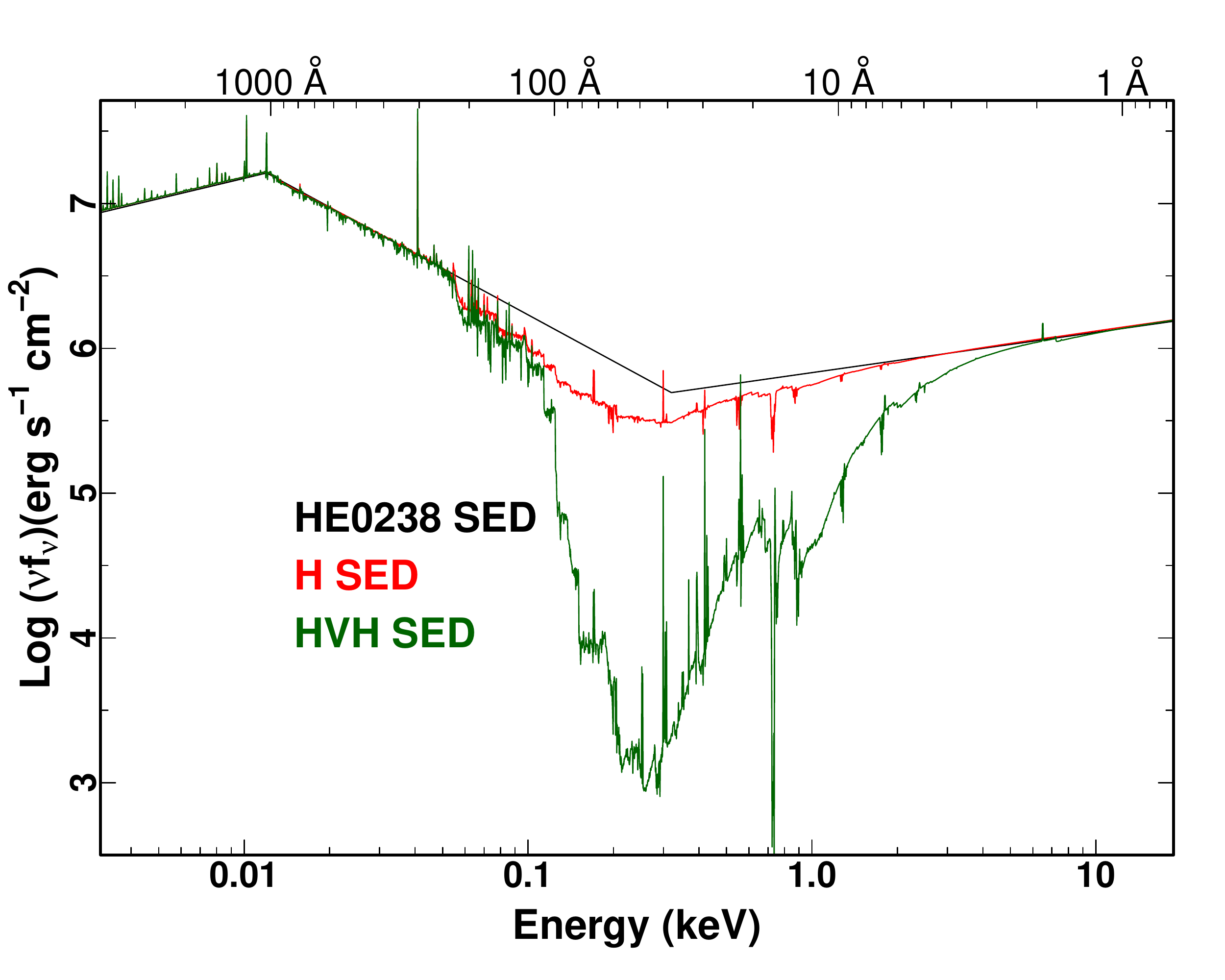}
	\caption{\footnotesize{Comparison between the HE0238 SED and the transmitted SEDs. The attenuations are similar in shape but stronger in absorption to what is seen in Paper II. \textit{Left Panel:} The VH SED (in blue) is the transmitted SED obtained by attenuating the HE0238 SED by the very high-phase solution of S2. The VHH SED in green is the transmitted SED obtained by having the HE0238 SED attenuated first by the very high-phase and then by the high-phase. \textit{Right Panel:} The H SED (in red) was obtained by attenuating the HE0238 SED by the high-phase solution of S2. The HVH SED in dark green is the transmitted SED obtained by having the HE0238 SED attenuated first by the high-phase and then by the very high-phase. }}
	\label{fig:sed}
\end{figure*}

Since S2 at 9~pc is interior to S1 at 500~pc, the SED seen by S1 will likely be attenuated by S2 \cite[e.g.,][]{bau10,sun17,mil18}. To test the effects this may have on the results of S1, we followed the procedure outlined in \cite{mil18}. We first generated new grids of Cloudy models using the transmitted SEDs from the high and very high photoionization solutions for S2 as well as the combination of the two (See Figure~\ref{fig:sed} for a comparison of each SED with the HE0238 SED). From these grids, new photoionization solutions and energetics were determined. For the shading from both the high-phase (H SED) and very high-phase (VH SED), $R$, $\dot{M}$, and $\dot{\sub{E}{K}}$ decreased by less than 30\%. The SEDs obtained by having the HE0238 SED attenuated first by the very high-phase and then by the high-phase (VHH SED) and vice versa (HVH SED) decreased $R$, $\dot{M}$, and $\dot{\sub{E}{K}}$ by about 50\%. These effects are small for the energetics, but the change in the distance is comparable to the error. However, the main conclusion that S1 is not contributing to major AGN feedback remains unchanged.
\section{Summary and Conclusions}
\label{sec:sc}
In this paper, we presented \textit{HST}/COS spectra for the quasar outflows seen in PKS J0352-0711. For the first time, we identified absorption troughs from ions \sion{Ca}{iv-v}, \sion{Ca}{v*}, and \sion{Ca}{vii*-viii*}. From the absorption troughs, ionic column densities for both outflow systems were calculated. A grid of photoionization models in conjunction with the ionic column densities enabled the determination of the best-fit solutions for \sub{U}{H} and \sub{N}{H} of each outflow system. 
%\\\\

The absorption troughs from \sion{O}{v*}, \sion{Ca}{vii*}, and \sion{Ca}{viii*} in S2 and \sion{O}{iv} and \sion{O}{iv*} in S1 yielded reliable density sensitive ratios. The \sion{O}{v*} column density ratio provided the S2 very high-phase \sub{n}{e}, and the \sion{Ca}{vii*} and \sion{Ca}{viii*} column density ratios independently confirmed the \sion{O}{v*} derived \sub{n}{e}. The \sion{O}{iv} density ratio yielded the S1 high-phase \sub{n}{e}. From these electron number densities, the distance to the central source of each outflow was calculated with equation (\ref{eq:R}). Equations (\ref{eq:M})~and~(\ref{eq:E}), along with the distance, enabled the determination of the mass flux and kinetic luminosity. Finally, the likely insignificant contribution each outflow has to AGN feedback processes was assessed, and these results were summarized in Table~\ref{tab:res}.
%\\\\

The following emerges from this work:
\begin{enumerate}
\item The extreme UV \textit{HST}/COS observations revealed never-before-seen absorption troughs with those from \sion{Ca}{vii*} and \sion{Ca}{viii*} being the most important. Their discoveries independently confirmed the \sion{O}{v*}--derived electron number density, distance, and energetics of S2.
\item Both outflow systems required a two-phase ionization solution, just like HE0238-1904, to satisfy the column density measurements of both the very high-ionization potential and high-ionization potential ions observed in each system.
\item The very high-ionization potential ions and the large associated hydrogen column density are similar to what is seen in the X-ray warm absorbers.
%\item Each outflow has a large enough Eddington ratio to be contributors to AGN feedback processes in PKS J0352-0711.
\end{enumerate}
%% If you wish to include an acknowledgments section in your paper,
%% separate it off from the body of the text using the \acknowledgments
%% command.

%% Included in this acknowledgments section are examples of the
%% AASTeX hypertext markup commands. Use \url without the optional [HREF]
%% argument when you want to print the url directly in the text. Otherwise,
%% use either \url or \anchor, with the HREF as the first argument and the
%% text to be printed in the second.

\acknowledgments

T.M., N.A., and X.X. acknowledge support from NASA  grants \textit{HST} GO-14777, 14242, 14054, and 14176. This support is provided by NASA through a grant from the Space Telescope Science Institute, which is operated by the Association of Universities for Research in Astronomy, Incorporated, under NASA contract NAS5-26555. T.M. and N.A. also acknowledge support from NASA ADAP 48020 and NSF grant AST 1413319. CHIANTI is a collaborative project involving George Mason University (USA), the University of Michigan (USA), and the University of Cambridge (UK). % G.L. acknowledges the grant from the National Key R\&D Program of China (2016YFA0400702), the National Natural Science Foundation of China (No. 11673020 and No. 11421303), and the National Thousand Young Talents Program of China. All authors are grateful to their home institutions for travel support, if provided, and to the anonymous referee whose careful review improved the quality of this paper.

\end{document}